\documentclass[11pt,a4paper]{article} 
\pdfoutput=1
\usepackage{adjustbox}
\usepackage{jheppub}
\usepackage{enumerate}
\usepackage{epsfig} 
\usepackage{contour}
\usepackage{tikz}
\usepackage{tikz-feynman}
\usepackage{float} 
\usepackage{subfigure}
\usepackage{wasysym} 
\usepackage{mathrsfs} 
\usepackage{amsfonts} 
\usepackage{amsbsy} 
\usepackage{amscd} 
\usepackage{pstricks} 
\usepackage{multirow} 
\usepackage{tikz}
\usepackage{color}
\usepackage{slashed}
\usepackage{array}
\usepackage{tablefootnote}
\usepackage{amsmath}
\usepackage{orcidlink}

\usetikzlibrary{arrows,positioning,shapes.geometric} 
\usepackage[font=small,labelfont=bf]{caption}
\usepackage{slashed}
\usepackage{multirow}
\usepackage{tabularx}
\usepackage{soul}  
\usepackage{listings}
\usepackage{color}

\usepackage{amsthm}

\newcommand{\pslash}{\slashed{q}}

\title{Impact of New Physics on Momentum-Dependent Particle Widths and Propagators}
\abstract{We investigate the impact of momentum-dependent particle widths and propagators on gauge and Higgs bosons and the top quark within the Standard Model (SM) and its SMEFT extensions near thresholds. By incorporating self-energy corrections via Dyson resummation, we quantify deviations from the fixed-width approximation and assess their implications for collider observables. While effects on the Higgs boson are negligible and the $W$ boson shows percent-level deviations in reconstructed transverse mass distributions, the top quark exhibits significant sensitivity near its mass threshold. Future lepton colliders, e.g., electron-positron machines or muon colliders, can offer sensitivity to these effects, enabling constraints on SMEFT Wilson coefficients. We perform a representative case study for the precision frontier available with a staged future muon collider. Our results highlight that momentum dependencies can provide additional sensitivity at precision-era experiments, enhancing the potential for discovering new physics there.}
\author[a]{Christoph Englert\orcidlink{0000-0003-2201-0667},}
\author[a]{Wrishik Naskar\orcidlink{0000-0002-4357-899},} 
\author[b]{and Michael Spannowsky\orcidlink{0000-0002-8362-0576}}
\affiliation[a]{School of Physics and Astronomy, University of Glasgow, Glasgow G12 8QQ, United Kingdom}
\affiliation[b]{Institute for Particle Physics Phenomenology, Department of Physics,\\ Durham University, Durham DH1 3LE, United Kingdom}
\emailAdd{christoph.englert@glasgow.ac.uk}
\emailAdd{w.naskar.1@research.gla.ac.uk}
\emailAdd{michael.spannowsky@durham.ac.uk}

\preprint{IPPP/25/02}

\allowdisplaybreaks

\begin{document}
\maketitle
\flushbottom

\section{Introduction}
A precise understanding of particle lifetimes is essential for accurately interpreting high-energy physics experiments that build their investigations around production cross sections and branching ratios of resonant ``signals'' compared to irreducible ``backgrounds''.  In the Standard Model (SM), fundamental particles such as the $W/Z$ bosons, the Higgs boson, and the top quark are subject to detailed scrutiny through their interactions. The intrinsic properties of these particles, such as their masses $m$, widths $\Gamma$ and propagators, then play a critical role in shaping observable distributions at collider experiments.

The finite lifetimes of particles in realistic quantum field theories such as the SM and its extensions are theoretically intriguing as they manifestly occur as higher-order effects in perturbation theory. Yet, they require careful treatment at the leading order to facilitate adequate comparisons of experiment and theory, avoiding drawbacks. For instance, the widely-used Breit-Wigner approach in this context typically implies gauge symmetry violation already at tree-level, see e.g.~\cite{Nowakowski:1993iu,Denner:2019vbn}.

Particle widths enter through higher-order effects, especially highlighted by applying the optical theorem~\cite{Lehmann:1954rq,1976opth,Cutkosky:1960sp} to the theory's 2-point functions. This links the imaginary part of the particles' self-energies $\Sigma(q^2)$ to the decay width as an on-shell quantity (with subtleties related to the presence of additional unstable particles in the spectrum~\cite{Veltman:1963th}). The full propagator of an unstable particle, in contrast to its leading order version, is therefore characterised by a pole in the second Riemann sheet of the S-matrix~\cite{Passarino:2010qk}, which, in turn, challenges textbook applications of the LSZ reduction (although solutions have been proposed~\cite{Weldon:1975gu}). In practice, careful consideration needs to be given to the definition of resonant ``signals'' that are naively associated with intermediate lines in Feynman diagrams to avoid gauge symmetry violation when moving beyond the tree approximation~\cite{Stuart:1991xk,Aeppli:1993rs,Seymour:1995np,Baur:1995aa}. Theoretical guidance presents itself through the application of Nielsen identities~\cite{Nielsen:1975fs,Gambino:1999ai,Grassi:2001bz,Goria:2011wa}, which enable a gauge-invariant definition of the complex resonance pole. Typically, particle widths are then consistently treated as input (renormalised mass) and derived (decay width) on-shell quantities in, e.g., in the complex mass scheme~\cite{Sirlin:1991fd,Nowakowski:1993iu,Denner:1999gp,Denner:2006ic,Actis:2006rc}.

A possible dependence of the width on momentum flow away from the resonance is not necessarily considered in particle phenomenology. This conventional approach greatly simplifies analyses, and it can be adequate when the expected experimental resolution is comparably low or the critical order parameter parametrising the relevance of such effects is small, i.e. $\Gamma/m\ll 1$. It may, however, overlook subtle effects that could be relevant, especially in the presence of new physics or higher-order corrections. When calculating the full propagator using Dyson-resummation, the resulting expression for the width includes the momentum-dependent terms from $\text{Im}[\Sigma(q^2)]$. This behaviour arises naturally as the decay phase space, available intermediate states, and loop corrections can vary with the energy or momentum of the decaying particle. It enters the Breit-Wigner propagator, e.g. for a bosonic particle, as (we will understand the Feynman prescription $i\epsilon$ as implicit in the following)
\begin{equation}
\label{eq:prop}
iG(q^2)=\frac{i}{q^2-m^2+im\Gamma(q^2)}\;,
\end{equation}
with $\Gamma(q^2) = \text{Im}[\Sigma(q^2)] / m$, modifying kinematic observables like the reconstructed mass of decaying resonances and cross sections of scattering processes.  Analyses of these effects, in particular as part of higher-order corrections and scheme dependencies have a long history in phenomenology~\cite{Wetzel:1982mh,Berends:1987ab,Beenakker:1988pv,Bardin:1988xt,Papavassiliou:1995fq,Passera:1996nk,Gambino:1999ai,Grassi:2001bz,Denner:2014zga}. Coincidentally, the heavy masses of the SM are traditionally extracted using the on-shell scheme, we will therefore employ this scheme throughout this work\footnote{Differences compared to the complex mass scheme can show scheme-dependencies much larger than the experimental uncertainties, e.g. in case of the $W$ and $Z$ bosons. Furthermore, in the top quark sector~\cite{Maltoni:2018zvp,Smith:1996xz}, additional systematic renormalon limitations~\cite{Beneke:1994sw,Bigi:1994em} are a recurring theme of topical conferences.}.

In this work, we aim to assess whether neglecting momentum dependence in particle widths is a valid simplification or whether it introduces discrepancies with phenomenological implications. By analysing momentum-dependent widths within the Standard Model Effective Field Theory (SMEFT) framework, we explore whether this commonly adopted simplification has measurable consequences or if the fixed-width approach remains sufficient within the aforementioned SM-related theoretical uncertainties. This could provide an opportunity to limit new physics, in particular in cases where the perturbative SM uncertainties are known to be small, e.g. in $Z$-pole analyses~\cite{Dittmaier:2009cr,Dittmaier:2014qza} or $t\bar t$ production~\cite{Kauer:2002sn}. Hence, we apply our analysis to various collider environments, including both hadron and lepton colliders, to evaluate the sensitivity of current and future experimental setups to potential SMEFT-induced deviations. 

This paper is organised as follows: We begin analysing the finite width effects for Standard Model particles in Section~\ref{sec:boson} for context. In Section~\ref{sec:wboson}, we analyse the $W$ boson propagator, detailing the derivation and implications of a momentum-dependent width. Section~\ref{sec:higgs} transitions to the Higgs boson, discussing its propagator and the effects of varying momentum on its width. In Section~\ref{sec:top}, we provide an in-depth analysis of the top quark propagator for the SM, discussing the perturbative calculations involved, where we can identify the potential for cornering physics beyond the SM scenarios. We then explore how SMEFT insertions modify the momentum-dependent top width, which assesses the implications of these findings at hadron and lepton colliders in Section~\ref{sec:topsmeft}. Finally, in Section~\ref{sec:conclusion}, we summarise our key findings and offer conclusions that underscore the broader impact of SMEFT-induced modifications on high-energy physics experiments. 

\section{Widths of SM particles}
\label{sec:boson}
We start by reviewing how widths naturally enter the propagators of unstable particles through higher-order effects, giving rise to poles in the second Riemann sheet of the S-matrix. This discussion introduces concepts and enables us to identify SM particles that are particularly susceptible to lineshape modifications from new physics. These modifications can then result in potential experimental consequences, particularly at the high-precision frontier of future colliders, which we will focus on in the next section.

\subsection{Higgs}
\label{sec:higgs}
We start by looking at the Higgs boson. Due to its scalar nature, the concepts become particularly transparent in this case. The renormalised Higgs one-particle irreducible two-point function is given by
\begin{equation*}
    \parbox{3.7cm}{\begin{tikzpicture}[baseline=(a)]
        \begin{feynman}
            \vertex (W1) at (-1, 0) {\(\large H\)};
            \vertex (a) at (0.1, 0);
            \vertex (b) at (0.9, 0);
            \vertex (W2) at (2, 0) {\(\large H\)};
            
            \diagram* {
                (W1) -- [scalar] (a),
                (b) -- [scalar] (W2)
            };
            \vertex[blob] (m) at (0.5,0) {};
        \end{feynman}
    \end{tikzpicture} }
    = \hat{\Gamma}^{H} (q^2) = -i (q^2 - m_H^2) - i \Sigma^H (q^2),
\end{equation*}
where $\Sigma^H $ denotes the self-energy of the Higgs boson considered at the one-loop level in the following. The on-shell renormalisation conditions for the Higgs two-point function $\hat{\Gamma}^H(q^2)$ are (see e.g.~\cite{Denner:1991kt})
\begin{equation}
    \begin{split}
        &\left.\text{Re}~\hat{\Gamma}^H (q^2)\right|_{q^2 = m_H^2} = 0, \\
        \lim_{q^2 \rightarrow m_H^2} &\left.\frac{1}{q^2-m_H^2}\text{Re}~\hat{\Gamma}^H (q^2)\right|_{q^2 = m_H^2} = i.
    \end{split}
\end{equation}
From the equations shown above, we obtain the following on-shell renormalisation conditions for the Higgs self-energy function $\Sigma^H (q^2)$
\begin{equation}
        \left.\text{Re}~\Sigma^H (q^2)\right|_{q^2=m_H^2} = \left.\text{Re}~\frac{\partial\Sigma^H (q^2)}{\partial q^2}\right|_{q^2=m_H^2} = 0.
\end{equation}
The Dyson-resummed Higgs propagator is then given in terms of the renormalised self-energy
\begin{subequations}
\label{eq:dyson}
\begin{equation}
    iG_H (q^2) = \frac{i}{q^2 - m_H^2 + \Sigma^H(q^2)}.
    \label{eqn:hprop}
 \end{equation}
By construction, this propagator exhibits unity pole residue when approaching the {\emph{real}} on-shell Higgs mass value, leaving an imaginary component that can be identified as a running width in comparison with~Eq.~\eqref{eq:prop}
\begin{equation}
    \Gamma_H (q^2) = \frac{\text{Im}~\Sigma^H(q^2)}{m_H}\quad\hbox{with}\quad\Gamma_H(q^2=m_H^2)\approx 5.5~\text{MeV},
\end{equation}
\end{subequations}
(for $m_H=125~\text{GeV}$ a the first relevant order) which is related to the tree-level computation of $H\to \{\text{SM~fields}\}$ due to the optical theorem (this value is also within the current limits, e.g. tabled by the PDG in~\cite{ParticleDataGroup:2024cfk}). 

\begin{figure}[!t]
    \centering
    \subfigure[\label{subfig:gammah}Higgs width.]{\includegraphics[width=0.54\linewidth]{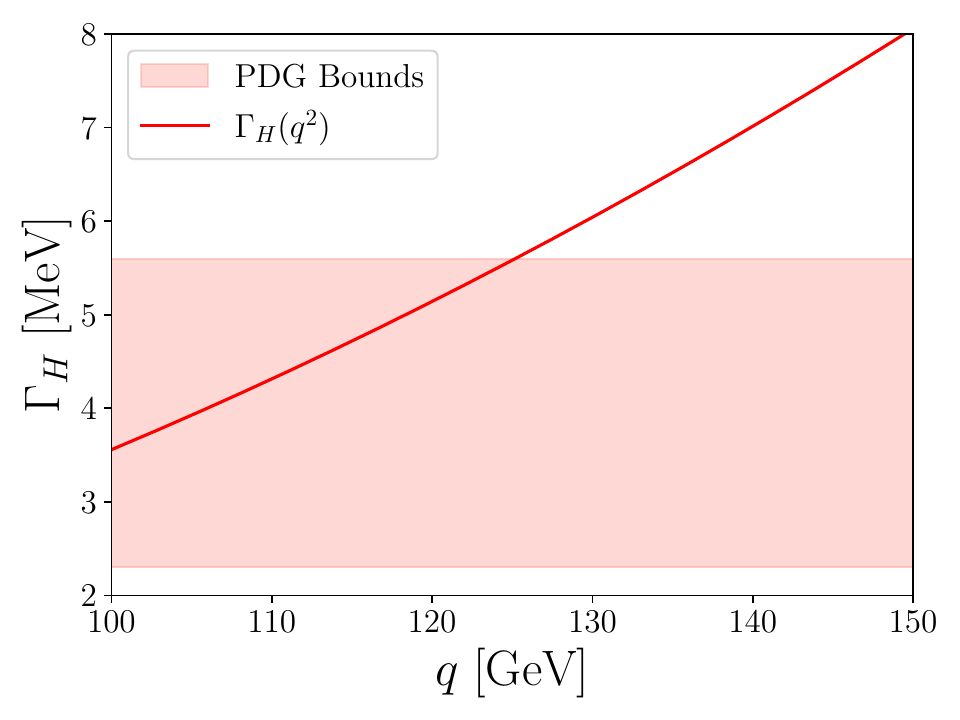}}\hfill
    \subfigure[\label{subfig:hprop}Higgs propagator.]{\includegraphics[width=0.42\linewidth]{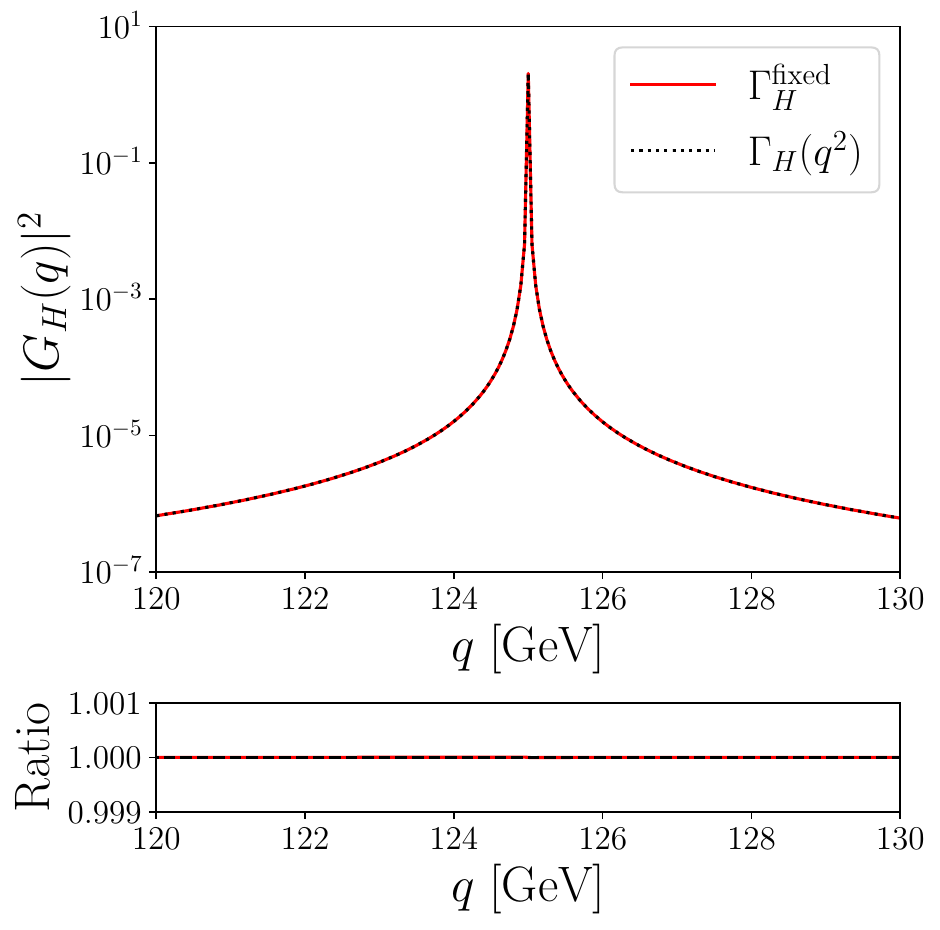}}
    \caption{Plots showing (a) the momentum-dependent Higgs width, and (b) the corresponding invariant masses of the Higgs using the propagators constructed with a fixed and a running width. The lower panel on the right shows the ratios between the two propagators.\label{fig:hplot}}
\end{figure}
Equation~\eqref{eq:dyson} means that close to the on-shell value of the Higgs boson, the loop-corrected propagator approximates the naive Breit-Wigner form, implying an exponential decay of the intermediately produced Higgs state from the squared matrix element. Representative plots illustrating the $q^2$ dependence of $\Gamma_H$ and the resulting $H$-propagator are shown in Fig.~\ref{fig:hplot} for the case of the SM. Given that the Higgs width is much smaller than its mass, the ratio of the two propagators shows minimal deviations from unity, visible in the lower panel of the plot in Fig.~\ref{subfig:hprop}. This clearly shows in parallel, that the ratio $\Gamma/m$ acts as a figure of merit for analyses that formulate results from pseudo-observable cross sections and branching ratios that effectively treat the intermediate particle (here the Higgs boson) as an asymptotic state of the S-matrix, bearing in mind scheme-dependencies. This also extends to concrete BSM scenarios, e.g.~\cite{Englert:2015zra,Kauer:2015hia}. It also becomes clear for a heavy SM Higgs (for a discussion before the Higgs discovery, see~\cite{Goria:2011wa}) as the Higgs mass parametrises the quality of the electroweak series' convergence, highlighted by unitarity constraints, which then particularly manifests itself in scheme dependencies in the treatment of the Higgs boson width as a quantity that is fundamentally sensitive to the distinction of different orders of the series expansion. In the SM, supported by current observations, we have $\Gamma^H/m_H \sim 10^{-5}$; any momentum-related modification of the Higgs on-shell region is therefore difficult to observe directly in the light of present and expected future resolution capabilities~\cite{deBlas:2019rxi,Dawson:2022zbb}, and is strongly constrained by existing pseudo-observable measurements. On the one hand, scheme dependencies, consequently, are numerically small~\cite{Zeng:2015gha,Zeng:2020lwi}. On the other hand, the influence of relevant BSM operators is predominantly constrained by modifications of these observables directly rather than from their momentum dependence, reshaping the resonance \cite{Englert:2014aca,Englert:2014ffa}. This makes the Higgs boson not a motivated candidate for the analysis that we are pursuing in this work.

\subsection{The $W$ boson}
\label{sec:wboson}
\begin{figure}[!t]
    \centering
    \includegraphics[width=0.6\linewidth]{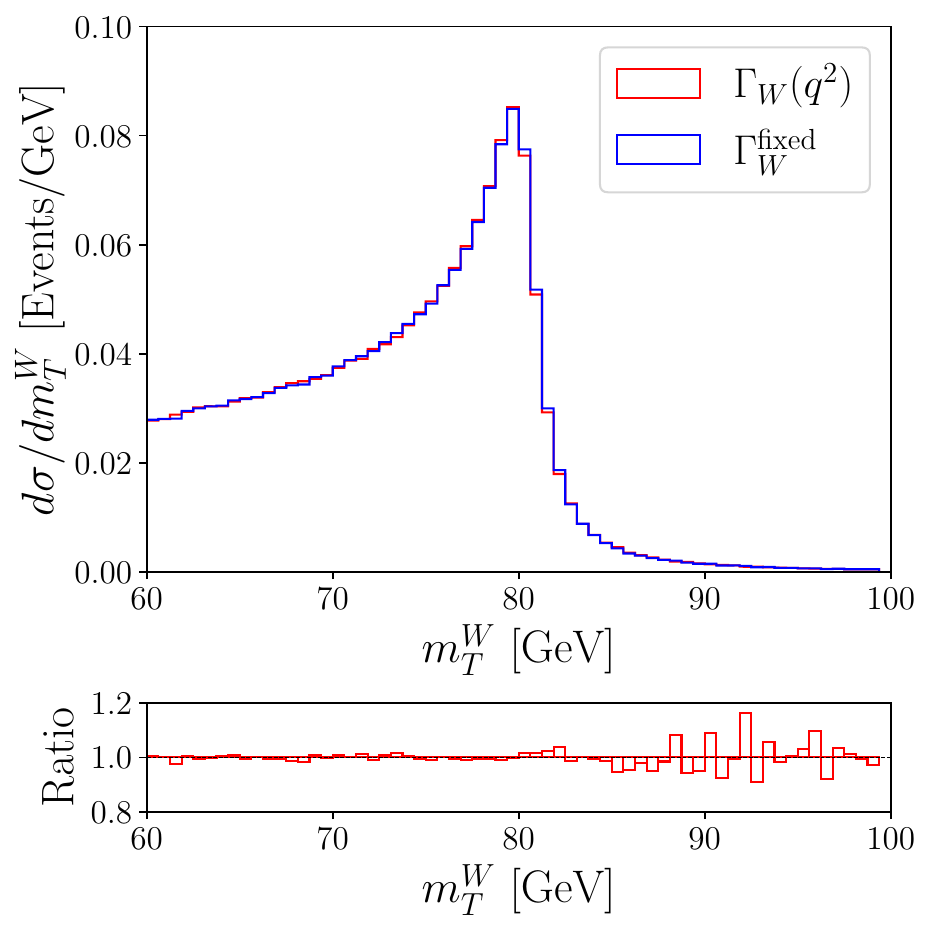}
    \caption{Differential distribution of the reconstructed transverse mass of the $W$ boson in 13~\text{TeV} $pp$ collisions comparing results for fixed and running widths. The lower panel shows the ratio of the two distributions.\label{fig:mtw}}
\end{figure}

Next, we consider the $W$ boson as a proxy for the massive gauge bosons. This enables us to be brief and transparent; the treatment of the $Z-\gamma$ mixing, which is detailed elsewhere in the literature, e.g.~\cite{Denner:1991kt}, yet the previous discussion of the Higgs boson generalises to all massive gauge bosons in broad terms. The renormalised two-point vertex function for the $W$ field can be written, already considering Feynman-'t Hooft gauge, as 
\begin{eqnarray}
    \parbox{3.9cm}{\begin{tikzpicture}[baseline=(a)]
        \begin{feynman}
            \vertex (W1) at (-1, 0) {\(\large W_\mu\)};
            \vertex (a) at (0.1, 0);
            \vertex (b) at (0.9, 0);
            \vertex (W2) at (2, 0) {\(\large W_\nu\)};
            
            \diagram* {
                (W1) -- [boson] (a),
                (b) -- [boson] (W2)
            };
            \vertex[blob] (m) at (0.5,0) {};
        \end{feynman}
    \end{tikzpicture} }
    &= &\hat{\Gamma}^W_{\mu \nu} (q)\nonumber\\
    &= &-i g_{\mu\nu}(q^2 - m_W^2) - i \left(g_{\mu\nu} - \frac{q_\mu q_\nu}{q^2}\right)\Sigma^{W}_T(q^2)-i\frac{q_\mu q_\nu}{q^2}\Sigma^{W}_L(q^2) ,\nonumber\\
\end{eqnarray}
and the renormalisation conditions for the $W$ two-point function for on-shell external physical fields are given by
\begin{equation}
    \begin{split}
        \left.\text{Re}~\hat{\Gamma}^W_{\mu \nu} (q) \varepsilon^\nu (q)\right|_{q^2 = m_W^2} &= 0,\\
        \lim_{q^2\rightarrow m_W^2}\left.\frac{1}{q^2-m_W^2}\text{Re}~\hat{\Gamma}^W_{\mu \nu} (q) \varepsilon^\nu (q)\right|_{q^2 = m_W^2} &= -i\varepsilon_\mu (q),
    \end{split}
\end{equation}
where $\varepsilon(q)$ is the polarisation vector of the external $W$ fields. Gauge-invariance and its manifestation in the gauge-fixed effective theory in terms of Slavnov-Taylor identities~\cite{Taylor:1971ff,Slavnov:1972fg} ensure that the longitudinal part of the propagator does not affect physical observables through the quartet mechanism~\cite{Kugo:1979gm}. Therefore, only the imaginary part of $\Sigma_W^T$ contributes directly to the decay width pseudo-observable. 
\begin{figure}[!t]
    \centering
    \subfigure[\label{subfig:gammaw}$W$ boson width.]{\includegraphics[width=0.54\linewidth]{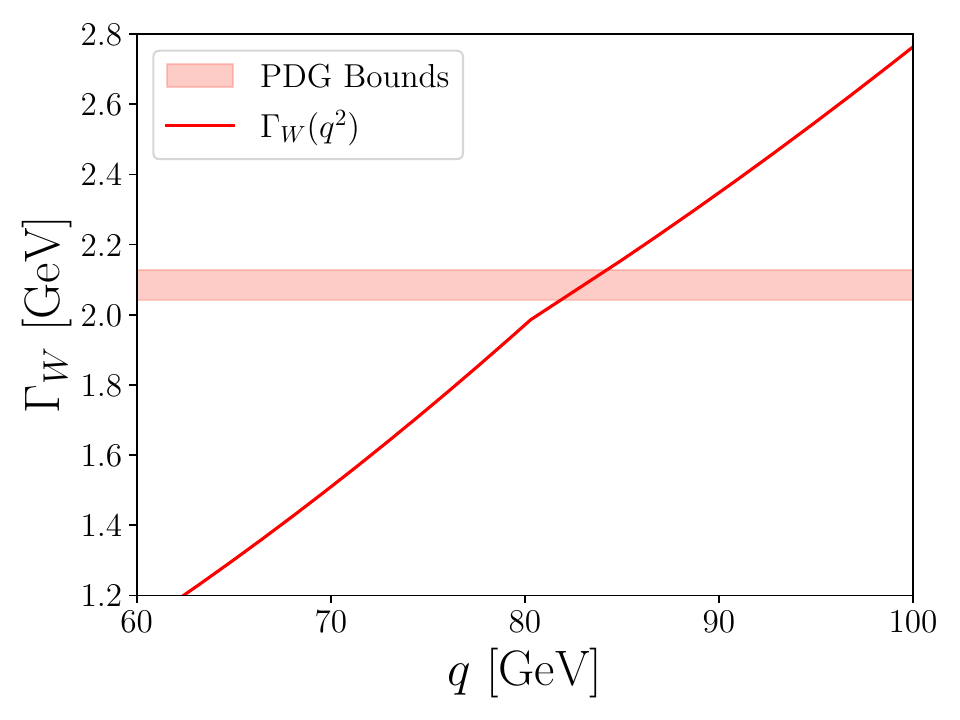}}\hfill
    \subfigure[\label{subfig:wprop}$W$ propagator.]{\includegraphics[width=0.42\linewidth]{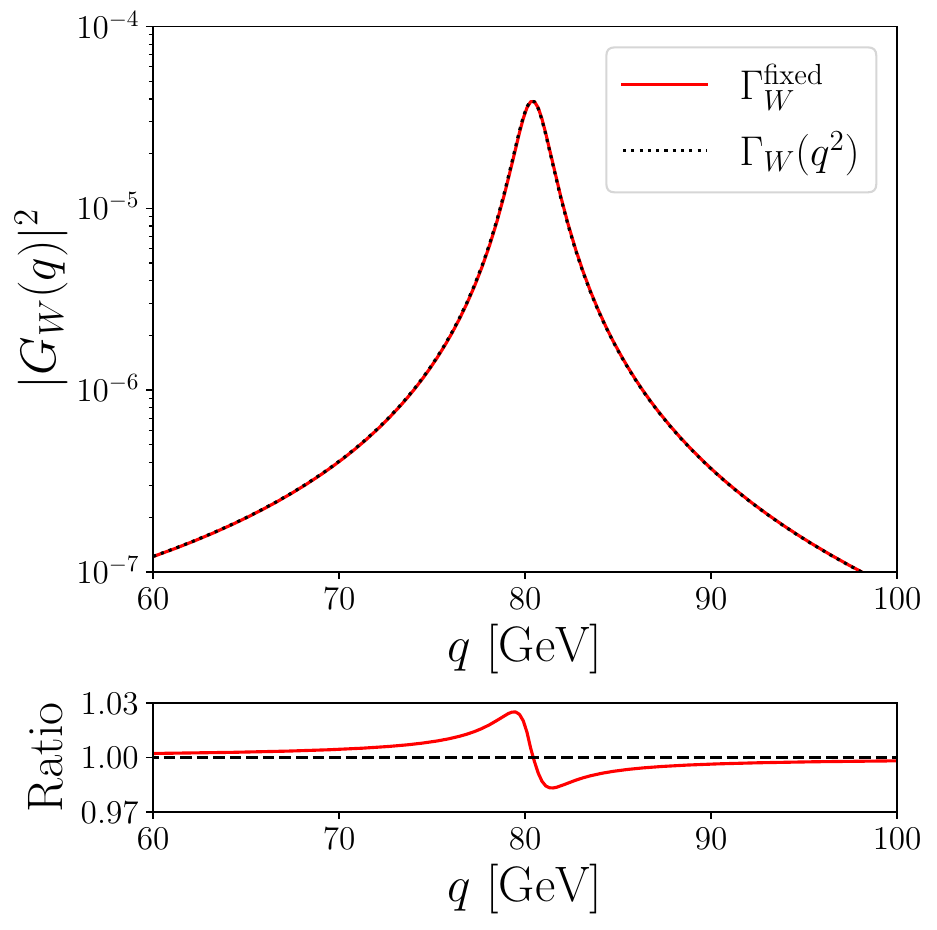}}
    \caption{Plots showing (a) the momentum dependent $W$-width, and (b) the corresponding invariant masses of the $W$ using the propagators constructed with a fixed and a running width. The lower panel shows the ratio between the respective propagators.
     \label{fig:wplot}}
\end{figure}
From the equations shown above, we obtain the following on-shell renormalisation conditions for the transverse part of the self-energy function
\begin{equation}
        \left.\widetilde{\text{Re}}~\Sigma_{T}^W (q^2)\right|_{q^2=m_W^2} = \left.\widetilde{\text{Re}}~\frac{\partial\Sigma_{T}^W (q^2)}{\partial q^2}\right|_{q^2=m_W^2} = 0.
\end{equation}
The Dyson-resummed $W$-propagator, involving only the renormalised transverse part $\Sigma^W_T$, is then given by 
\begin{equation}
   \label{eqn:wprop}
   i G_W^{\mu\nu} (q^2) =  \frac{-i}{q^2 - m_W^2 + \Sigma^W_T(q^2)} \left( g^{\mu \nu} - \frac{q^\mu q^\nu}{q^2} \right).
\end{equation}
From the propagator shown in Eq.~\eqref{eqn:wprop}, the $q^2$-dependent running width of the $W$ boson can be identified as
\begin{equation}
    \Gamma_W (q^2) = \frac{\text{Im}~\Sigma_T^W(q^2)}{m_W}\quad \text{with}\quad\Gamma_W (q^2 = m_W^2)\approx 1.99~\text{GeV}
\end{equation}
similar to the Higgs boson. The dependence of $\Gamma_W$ on $q^2$ and the corresponding $W$ propagator are shown in Fig.~\ref{fig:wplot}. We observe variations at the percent level in the $W$ boson invariant mass lineshape as illustrated in Fig.~\ref{subfig:wprop}. It is of interest however, to explore whether the running width has implications for the precise $W$-mass measurements reported by CDF~\cite{CDF:2022hxs}, ATLAS~\cite{ATLAS:2024erm}, and very recently CMS~\cite{CMS-PAS-SMP-23-002}, where each analysis measured the reconstructed transverse mass of the $W$ ($m_W^T$). By following the methodology in Ref.~\cite{CDF:2022hxs}, we obtain the differential $m_W^T$ distribution for both fixed and running $\Gamma_W$, as shown in Fig.~\ref{fig:mtw}. As expected, the percent level differences in the W-propagator do not offer significant deviations in the transverse mass differential distribution. Of course, this momentum dependence should be viewed against the complete set of electroweak corrections, which are scheme and process-dependent. The conclusion for us at this point is that the $W$ propagation does not provide significant a priori potential to constrain the momentum dependencies away from the peak observables, resulting in a subleading effect for phenomenological analyses \cite{Banerjee:2024eyo}.
It can be checked that the momentum-dependent width of the $Z$ boson exhibits a behaviour similar to that of the $W$ boson, with quantitatively distinct contributions from fermion loops. Despite these differences, the variations in the momentum-dependent width of the $Z$ boson are even smaller than those observed for the $W$ boson. Consequently, any potential impacts on the invariant mass or transverse mass distributions of the $Z$ boson are negligible, offering even reduced potential compared to our discussion of the $W$ boson.
\subsection{The Top Quark}
\label{sec:top}
Finally, we extend our analysis of calculating the momentum-dependent width to the top quark. As for the discussion of the previous sections, the study of the top resonance characteristics provides valuable insights into possible BSM effects. Again most analyses so far have relied on peak pseudo-observables (e.g. ATLAS analyses like~\cite{ATLAS:2023gsl} used in the EFT analyses like \cite{Buckley:2015lku,Hartland:2019bjb,Brivio:2019ius,Bissmann:2020mfi,Ethier:2021bye,Garosi:2023yxg,Atkinson:2024hqp}). Recently, there has been a resurgence in the literature on hypothetical bound states of top and antitop quarks, \textit{toponium}~\cite{Aguilar-Saavedra:2024mnm,Fuks:2024yjj,Garzelli:2024uhe} triggered by anomalies of LHC data. While toponium itself remains a short-lived entity due to the rapid decay of the top quark (\emph{before} the bound state is formed), it can alter distributions at threshold production in, e.g., top pair production. Therefore, this currently evolving situation can be considered an additional source of uncertainty for the topics discussed in this section.

\begin{figure}[!t]
    \centering
    \includegraphics[width=0.7\linewidth]{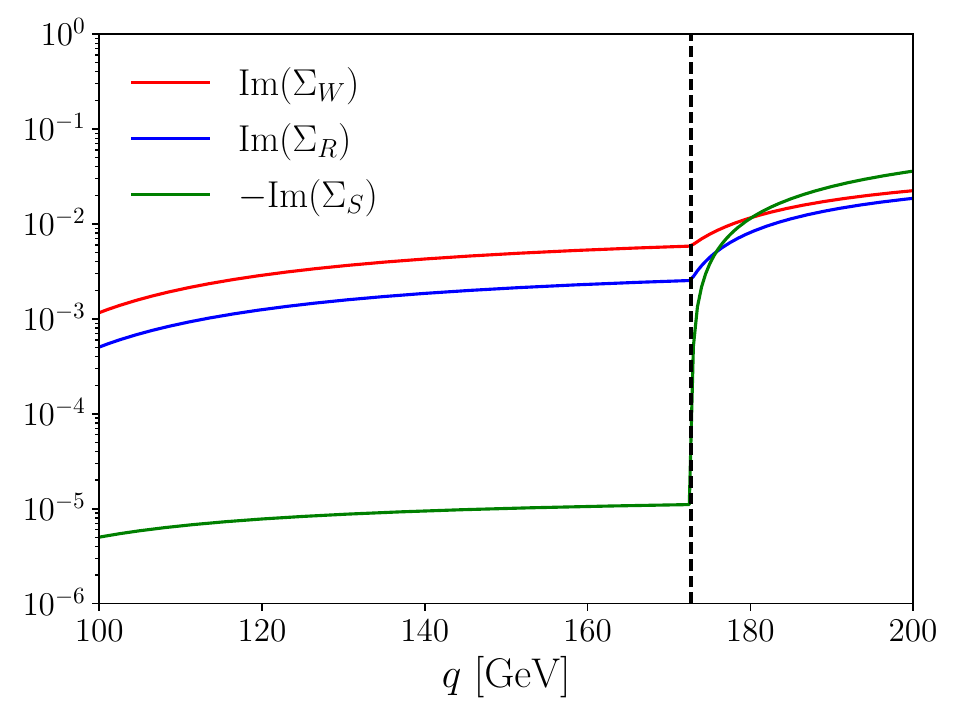}
    \caption{The imaginary parts of the different two-point amplitudes associated with the top quark self-energy.\label{fig:tsigs}}
\end{figure}

\begin{figure}[!t]
    \centering
    \subfigure[\label{subfig:gammat}Top width as detailed in the text.]{\includegraphics[height=0.39\linewidth]{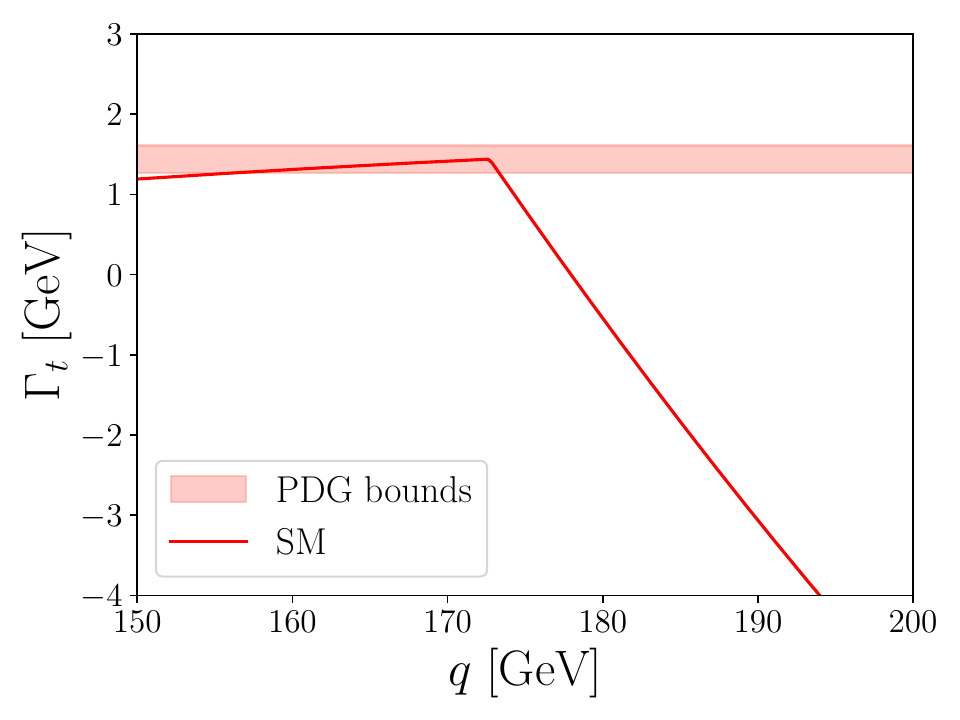}}\hfill
    \subfigure[\label{subfig:tprop}Top propagator for different approximations.]{\includegraphics[height=0.39\linewidth]{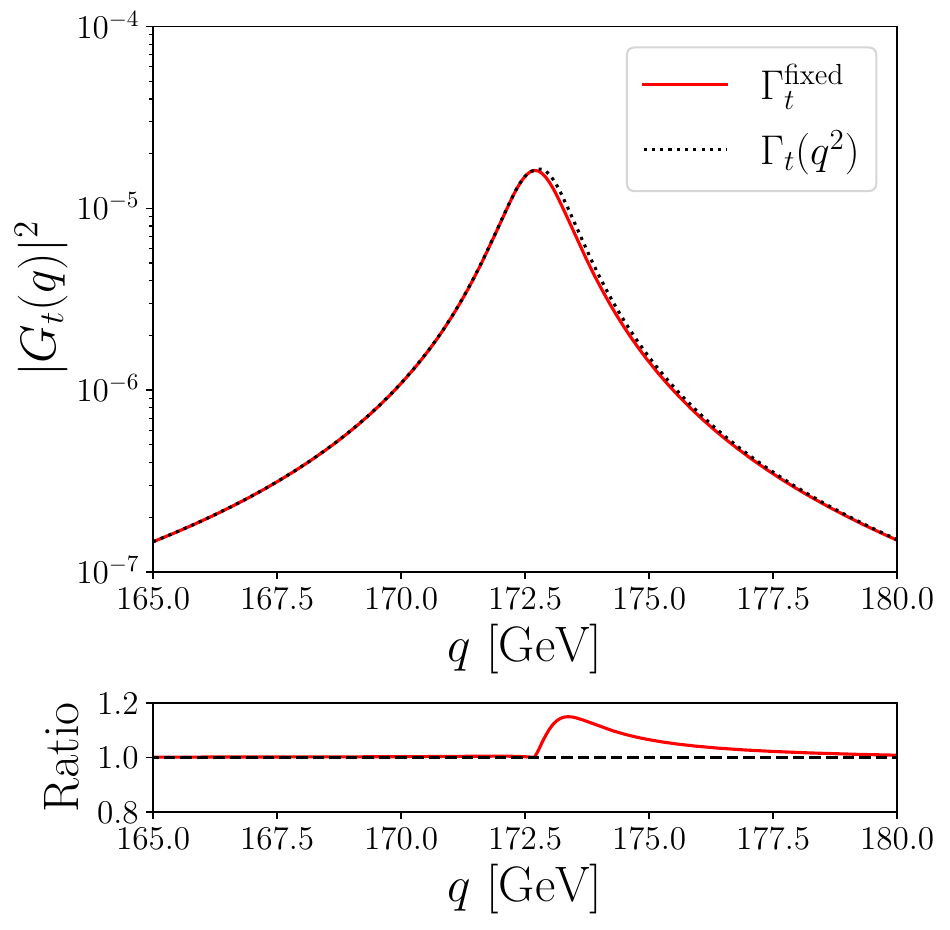}}
    \caption{Plots showing (a) the momentum-dependent top width and (b) the corresponding invariant masses of the top using the propagators constructed with a fixed and a running width (right). The lower panel on the right represents the ratio of the two plotted propagators.\label{fig:tplot}}
\end{figure}

Building upon the framework established in the previous sections for the $W$ boson and the Higgs, we delve into the unique features of the top quark width and the corresponding propagator, whose calculations are comparatively more involved than the previous cases. We begin by looking at the renormalised 1PI two-point function for the top quark,
\begin{equation*}
    \parbox{4.4cm}{\begin{tikzpicture}[baseline=(a)]
        \begin{feynman}
            \diagram[horizontal=a to b, large] {
                i1 [particle=\(t\)] -- [fermion] a -- b[blob] -- c -- [fermion] i2 [particle=\(t\)],
            };
        \end{feynman}
    \end{tikzpicture} }
    = i(\pslash - m_t) + i \left[\pslash \omega_{-} \Sigma_{L}^t (q^2) + \pslash \omega_{+} \Sigma_{R}^t (q^2) + m_t \Sigma_{S}^t (q^2)\right],
\end{equation*}
where $\omega_{\pm} = (1 \pm \gamma^5)/2$ are the chiral projectors. The on-shell renormalisation conditions for the two-point functions read
\begin{equation*}
    \begin{split}
        \left.\text{Re}~\Sigma_{L}^t (q^2)\right|_{q^2 = m_t^2} &+ \left.\text{Re}~\Sigma_{S}^t (q^2)\right|_{q^2 = m_t^2} = 0, \\
        \left.\text{Re}~\Sigma_{R}^t (q^2)\right|_{q^2 = m_t^2} &+ \left.\text{Re}~\Sigma_{S}^t (q^2)\right|_{q^2 = m_t^2} = 0, \\
        \left.\text{Re}~\Sigma_{L}^t (q^2)\right|_{q^2 = m_t^2} &+ \left.\text{Re}~\Sigma_{R}^t (q^2)\right|_{q^2 = m_t^2} \\
        + 2 m_t^2 & \left.\frac{\partial}{\partial q^2}\left[\text{Re}~\Sigma_{L}^t (q^2) + \text{Re}~\Sigma_{R}^t (q^2) + 2 \text{Re}~\Sigma_{S}^t (q^2)\right]\right|_{q^2=m_t^2} = 0.
    \end{split}
\end{equation*}
The plot of the imaginary parts of $\Sigma_L$, $\Sigma_R$, and $\Sigma_S$ that are relevant for the top width calculations in the SM are depicted in Fig.~\ref{fig:tsigs}. Defining 
\begin{equation*}
i\Sigma_2^t (\pslash,q^2) = i \left[\pslash \omega_{-} \Sigma_{L}^t (q^2) + \pslash \omega_{+} \Sigma_{R}^t (q^2) + m_t \Sigma_{S}^t (q^2)\right],
\end{equation*}
we obtain the full top propagator, including the sum of all 1PI insertions. 
\begin{equation*}
    \begin{split}
    i G(\pslash,q^2) &=  
    \parbox{1.45cm}{\begin{tikzpicture}[baseline=(a)]
        \begin{feynman}
            \diagram[horizontal=a to b, small] {
                a [particle=\(t\)] -- [fermion] b [particle=\(t\)],
            };
        \end{feynman}
    \end{tikzpicture}} + 
     \parbox{2.45cm}{\begin{tikzpicture}[baseline=(a)]
        \begin{feynman}
            \diagram[horizontal=a to b, small] {
                i1 [particle=\(t\)] -- [fermion] a b[blob] -- [fermion] i2 [particle=\(t\)]
            };
        \end{feynman}
    \end{tikzpicture}} + 
     \parbox{3.45cm}{\begin{tikzpicture}[baseline=(a)]
        \begin{feynman}
            \diagram[horizontal=a to b, small] {
                i1 [particle=\(t\)] -- [fermion] a b[blob] -- [fermion] c[blob] -- [fermion] i2 [particle=\(t\)],
            };
        \end{feynman}
    \end{tikzpicture}} + ...\\
    &\hspace{-1.5cm}= \frac{i}{\pslash - m_t} + \left[\frac{i}{\pslash - m_t} i \Sigma_2^t (\pslash,q^2) \frac{i}{\pslash - m_t}\right] + \left[\frac{i}{\pslash - m_t} i \Sigma_2^t (\pslash,q^2) \frac{i}{\pslash - m_t} i \Sigma_2^t (\pslash,q^2) \frac{i}{\pslash - m_t}\right] + ...\\
    &= \frac{i}{\pslash - m_t + \Sigma_2^t (\pslash,q^2)}
    \end{split}
\end{equation*}
The full expression for the top quark propagator is
\begin{equation}
    i G(\pslash,q^2) = \frac{i}{\pslash [1 + \omega_{-} \Sigma_{L}^t (q^2) + \omega_{+} \Sigma_{R}^t (q^2)] - m_t[1 - \Sigma_{S}^t (q^2)]}.
    \label{eq:tprop}
\end{equation}
The Breit-Wigner form then follows from Eq.~\eqref{eq:tprop} as
\begin{equation}
\label{eq:BW2}
        i G(\pslash,q^2) = i \frac{\pslash [1 + \omega_{-} \Sigma_{L}^t (q^2) + \omega_{+} \Sigma_{R}^t (q^2)] + m_t[1 - \Sigma_{S}^t(q^2)]}{q^2[1 + \Sigma_{L}^t(q^2)][1+\Sigma_{R}^t(q^2)] - m_t^2 [1 - \Sigma_{S}^t (q^2)]^2}.
\end{equation}
Starting from our result in Eq.~\eqref{eq:BW2}, isolating the $q^2$ term in the denominator, the propagator takes the form,
\begin{equation}
    i G(\pslash,q^2) = i \frac{\pslash [1 + \omega_{-} \Sigma_{L}^t (q^2) + \omega_{+} \Sigma_{R}^t (q^2)] + m_t[1 - \Sigma_{S}^t(q^2)]}{(1 + \Sigma_{L}^t(q^2))(1+\Sigma_{R}^t(q^2))\left(q^2 - \frac{m_t^2 [1 - \Sigma_{S}^t (q^2)]^2}{[1 + \Sigma_{L}^t(q^2)][1+\Sigma_{R}^t(q^2)]}\right)},\label{eq:BWt2}
\end{equation}
We can then express, see e.g.~\cite{Dreiner:2008tw},
\begin{equation}
    M_{t,\text{pole}}^2 - i M_{t,\text{pole}} \Gamma_t = \left.\frac{m_t^2 (1 - \Sigma_S)^2}{(1 + \Sigma_L)(1 + \Sigma_R)}\right|_{q^2=m_t^2}.
    \label{eq:BWt3}
\end{equation}
To one-loop order, Eq.~\eqref{eq:BWt3} can be expanded perturbatively
\begin{equation*}
    M_{t,\text{pole}}^2 - i M_{t,\text{pole}} \Gamma_t = \left.m_t^2\left(1 - \Sigma_L - \Sigma_R - 2 \Sigma_S\right)\right|_{q^2=m_t^2}.
\end{equation*}
From this the pole-mass and the width can then be derived~\cite{Dreiner:2008tw} (for $m_t = 172.7~\text{GeV}$~\cite{ParticleDataGroup:2024cfk}) 
\begin{equation}
    \begin{split}
        M_{t,\text{pole}} &= m_t \left.\left(1 - \frac{1}{2}\Sigma_L - \frac{1}{2}\Sigma_R -\Sigma_S\right)\right|_{q^2=m_t^2}\approx 172.7~\text{GeV}\\
        \Gamma_t(q^2) &= m_t~\text{Im} \left(\Sigma_L + \Sigma_R + 2 \Sigma_S\right),\quad \Gamma_t(q^2 = m_t^2)\approx 1.44~\text{GeV}.
    \end{split}\label{eq:wid3}
\end{equation}
This result is consistent with the current PDG value for $\Gamma_t$~\cite{ParticleDataGroup:2024cfk}. In the narrow-width approximation, where $\Gamma_t \ll m_t$, the top quark propagator can be written up to one-loop order as 
\begin{equation}
    i G^t(\pslash,q^2) = i\frac{\pslash + m_t}{q^2 - m_t^2 + i m_t^2 \text{Im} \left(\Sigma_L + \Sigma_R + 2 \Sigma_S\right) }.
\end{equation}
The $q^2$-dependent running width described by Eq.~\eqref{eq:wid3} and the corresponding propagators constructed from the fixed and the running top width have been plotted in Fig.~\ref{fig:tplot}. We observe a pronounced change in the running width after $q^2 = m_t^2$, discernible from the two-point amplitudes characterised by $\Sigma_L,~\Sigma_R$ and notably $\Sigma_S$ as shown in Fig.~\ref{fig:tsigs}. The marked increase in the imaginary parts of these amplitudes at $q^2 = m_t^2$ arises due to the top quark self-energy contributions, specifically from the gluonic diagrams turning on at $m_t$ due to the optical theorem. This behaviour is further reflected in the top propagator as depicted in Fig.~\ref{fig:tplot}, where we observe a significant deviation from the propagator constructed with the fixed width, as illustrated in the lower panel on the right. Thus, any new physics scenarios that modify the top quark two-point functions may leave a substantial imprint on the running width and, consequently, on the invariant mass, which may be detectable by the extensive array of current and future colliders. It should again be noted that these effects will be part of a larger set of electroweak corrections. Especially when dealing with the massless fields that shape distributions as detailed above, the matching of infrared singularities becomes relevant, see e.g.~\cite{Dittmaier:2014qza} for a detailed discussion. In this context, however, it is worth stressing that, by construction, EFT-deformations do not modify the soft and collinear behaviour of the SM~\cite{Englert:2018byk}. So, while the presence of virtual massless states requires care from the point of view of the dimension-four interactions, the relation of dimension-six interactions to the SM is not sensitive to such an interplay. However, the emission of hard photons or gluons certainly poses a direct avenue to limit interaction modifications. We will comment on this relation at the end of the next section.

\section{New Physics effects in Running Top Quark Widths and Propagators}
\label{sec:topsmeft}
\begin{table}[!b]
	\centering
	\renewcommand{\arraystretch}{1.7}	
    \begin{tabular}{|c|c||c|c|}
			\hline 			 
			$\mathcal{O}_{tG}$ & $(\bar{t}_L \sigma^{\mu\nu} T^A t_R) \tilde{\phi} G^{A}_{\mu \nu}$  
            & $\mathcal{O}_{tW}$ & $(\bar{t}_L \sigma^{\mu\nu} t_R) \tau^I \tilde{\phi} W^{I}_{\mu \nu}$
			\\
			$\mathcal{O}_{t\phi}$ & $(\phi^\dagger \phi) (\bar{t}_L t_R \phi)$
			& $\mathcal{O}_{tB}$ & $(\bar{t}_L \sigma^{\mu\nu} t_R) \tilde{\phi} B_{\mu \nu}$
			\\
			\hline			
	\end{tabular}
	\caption{Dimension-six SMEFT operators~\cite{Grzadkowski:2010es} in the top quark sector that modify the top quark two-point function. $\tilde \phi$ is the charge conjugated SM Higgs doublet.} 
	\label{tab:ops_top}
\end{table}
We use the Standard Model Effective Field Theory (SMEFT) framework to examine the effects of new physics on the $q^2$-dependent running width and the resulting propagator of the top quark. We modify the top two-point function by introducing dimension-six SMEFT operators in the Warsaw basis~\cite{Grzadkowski:2010es}, which alter the top quark vertices that appear in self-energy diagrams. The SMEFT operators considered in our analysis are listed in Tab.~\ref{tab:ops_top}, all in the top quark sector. The corresponding Wilson coefficients hold significant BSM potential due to the currently relatively loose constraints on them ($\mathcal{O}(0.1-1)~\text{TeV}^{-2}$~\cite{Garosi:2023yxg,Celada:2024mcf}). Firstly, we only consider the SMEFT operator $\mathcal{O}_{tG}$, which modifies the $ttg$ vertex to capture the effects of the gluonic topologies contributing to the top self-energy. This operator, therefore, does not affect $\Gamma_t (q^2)$ directly, contributing solely to the tails of the top invariant mass lineshape. However, this scenario changes with the inclusion of additional effective operators, specifically $\mathcal{O}_{tW}$, $\mathcal{O}_{tB}$, and $\mathcal{O}_{t\phi}$. A notable subtlety arises here, as the operator $\mathcal{O}_{tW}$ directly modifies $\Gamma_t (t \rightarrow W b)$; any phenomenologically relevant momentum dependence therefore could therefore enhance the new physics potential, but unlikely drive it. 

\begin{figure}[!t]
    \centering
    \subfigure[\label{subfig:gammateft}Modified top width.]{\includegraphics[width=0.49\linewidth]{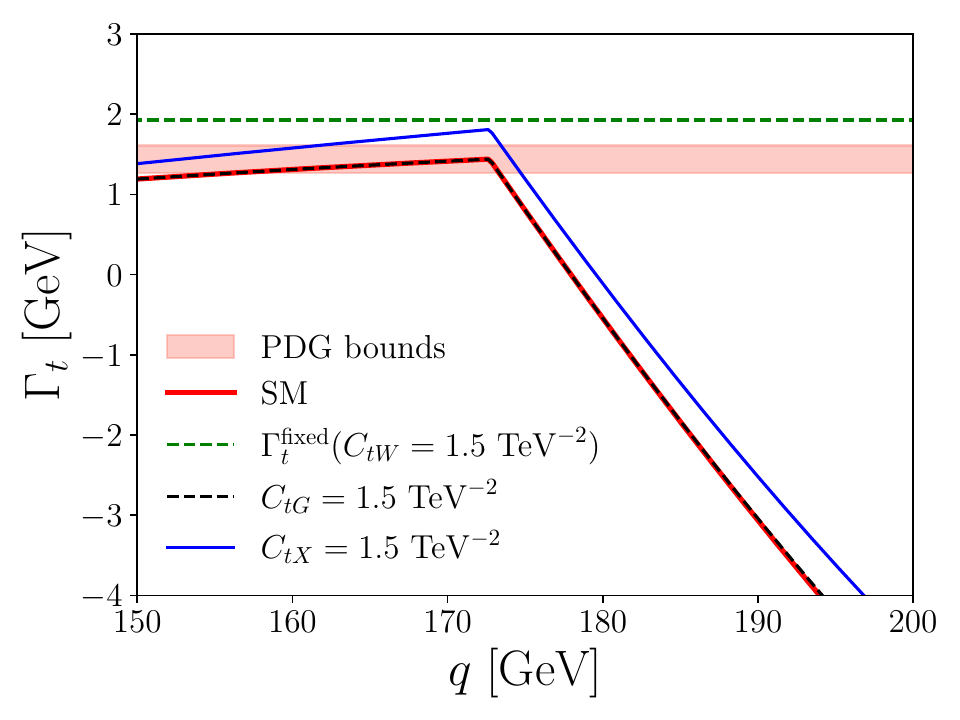}}\hfill
    \subfigure[\label{subfig:tpropeft}Modified top propagator.]{\includegraphics[width=0.49\linewidth]{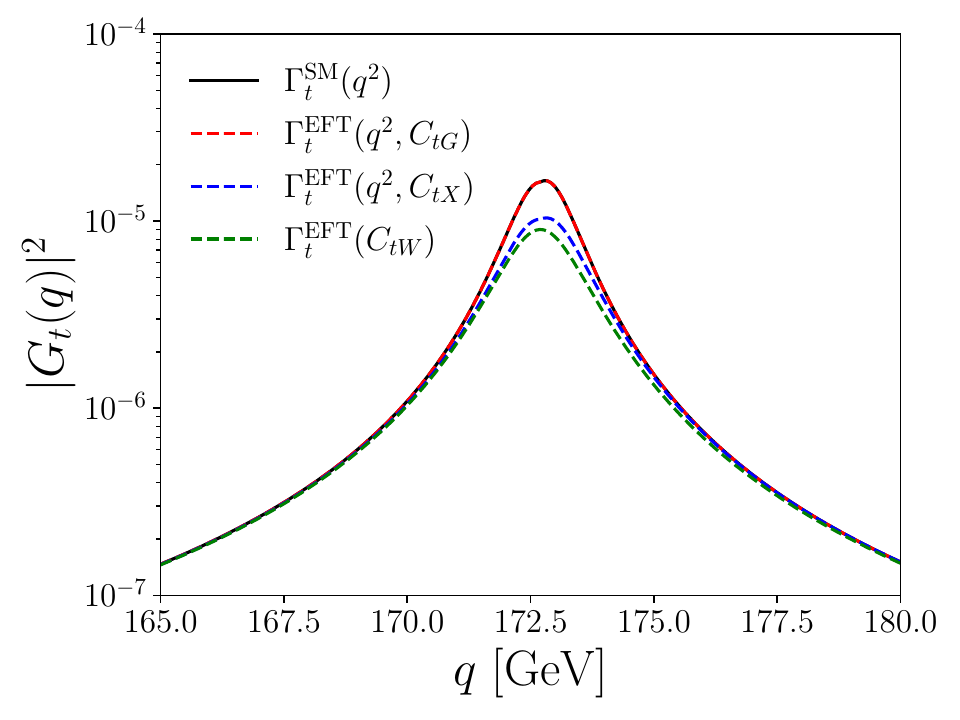}}
    \caption{Plots showing (a) the momentum-dependent top width, and (b) the corresponding invariant masses of the top using the propagators constructed with a running width for different SMEFT insertions, and a fixed width corresponding to a direct modification of $t \rightarrow W^+ b$ decay. In both plots, $C_{tW} = C_{tB} = C_{tG} = C_{t\Phi} = C_{tX} = 1.5~\text{TeV}^{-2}$.\label{fig:tploteft}}
\end{figure}

We follow the same renormalisation procedure as described in the previous section to derive expressions for the perturbative width up to one-loop order 
\begin{equation*}
\Gamma_t = \text{Im}\left[m_t(\Sigma_L + \Sigma_R + 2\Sigma_S)\right].
\end{equation*} 
The $q^2$-dependent widths and the corresponding propagators for both scenarios are presented in Fig.~\ref{fig:tploteft}. Additionally, we depict the fixed width and the corresponding propagator obtained by examining the direct contribution of $\mathcal{O}_{tW}$ on the $t \rightarrow W b$ decay. As anticipated, inserting only $\mathcal{O}_{tG}$ does not alter the width calculated at $q^2 = m_t^2$. However, by observing the running width at higher $q^2$, a slight divergence from the Standard Model running width is apparent for the selected value of the associated Wilson coefficient. The running width computed with multiple operator insertions, assuming equal values for the relevant Wilson coefficients, deviates significantly from the Standard Model running width and the modifications induced by the direct alteration of $t \rightarrow W b$.
\begin{figure}[!t]
    \centering
    \includegraphics[width=0.6\linewidth]{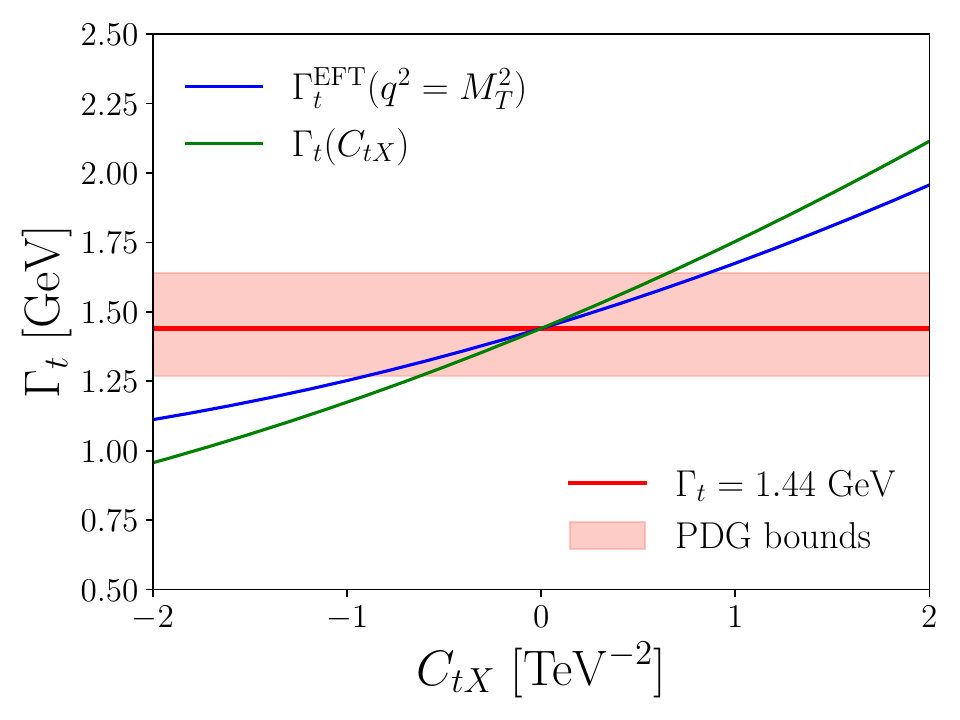}
    \caption{Wilson coefficient dependence of $\Gamma_t$ computed from (1) running width with SMEFT insertions at $q^2 = m_t^2$, and (2) from $t \rightarrow W^+ b$. Here, $C_{tW} = C_{tB} = C_{tG} = C_{t\Phi} = C_{tX}$.\label{fig:tploteftmod}}
\end{figure}
This difference can also be visualised by examining the dependence of the running width on the Wilson coefficients at $q^2 = m_t^2$ and contrasting this with the modification of the fixed decay width, as illustrated in Fig.~\ref{fig:tploteftmod}. Given that new physics affecting the Wilson coefficients under consideration may lead to observable changes in the running width of the top quark, it is crucial to explore whether observables sensitive to running width effects could provide constraints on these Wilson coefficients comparable to those achieved through global fit methodologies. This will be the focus of the subsequent sections.
\subsection{Hadron Colliders}
\begin{figure}[!t]
    \centering
    \includegraphics[width=0.6\linewidth]{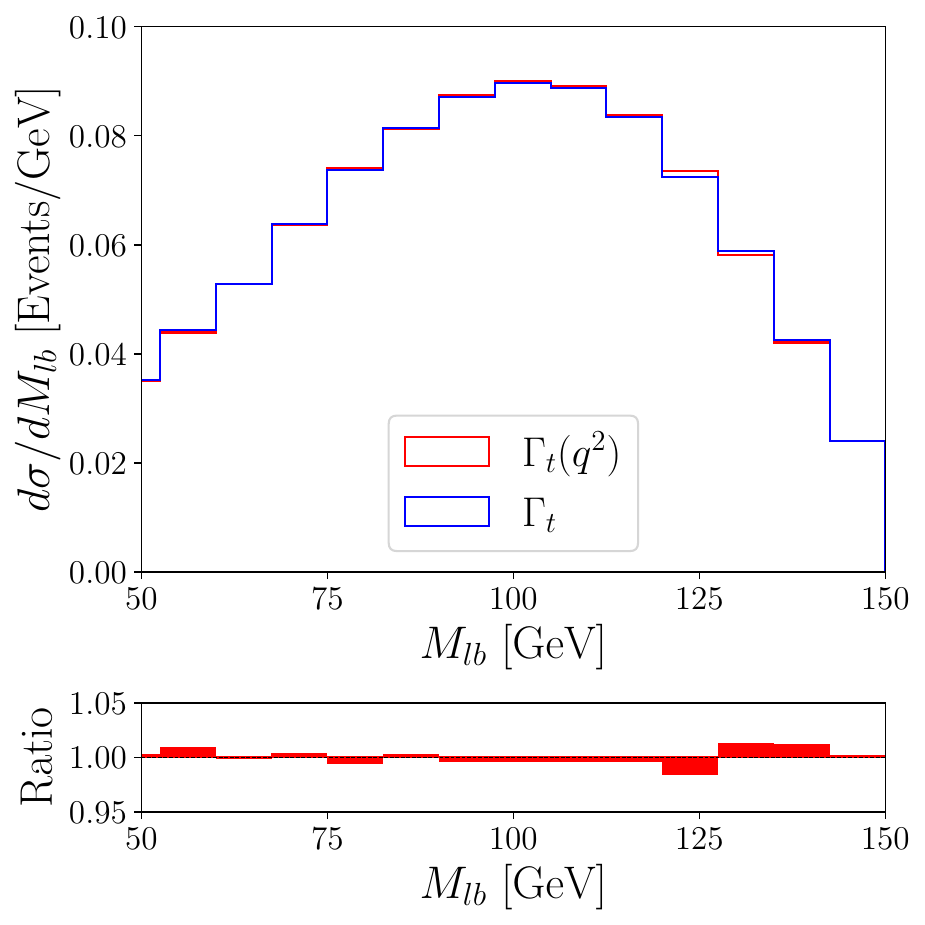}
    \caption{Differential distribution of the reconstructed invariant mass of a lepton and the corresponding $b$-jet ($M_{lb}$) in the dileptonic decay of $t\bar{t}$ production in $13~\text{TeV}$ $pp$-collisions, for fixed and running top widths. 
    \label{fig:mlb}}
\end{figure}

For instance, the ATLAS collaboration has conducted a direct measurement of the top quark decay width~\cite{ATLAS:2019onj} at $\sqrt{s} = 13$~TeV, with an integrated luminosity of $\mathcal{L} = 139~\text{fb}^{-1}$. This measurement utilised a profile-likelihood template fit applied to the invariant mass distribution of a charged lepton and its associated $b$-jet ($M_{lb}$) in the dileptonic decay channel of $t\bar{t}$-production. Consequently, it is essential to employ their search strategy to assess the impact of running widths on the kinematic properties of the top quark. Following the approach outlined in Ref.~\cite{ATLAS:2019onj}, we derive differential distributions for the kinematic observable $M_{lb}$, comparing the Standard Model fixed and running top quark widths. Our findings, displayed in Fig.~\ref{fig:mlb}, indicate that hadron colliders exhibit limited sensitivity (variations only at the percent level) to running width effects of the top quark, feeding into a low BSM potential along the lines we pursue in this work.
\subsection{Lepton Colliders}
\label{subsec:lepcol}
Unlike hadron machines, future lepton colliders offer unique advantages by providing cleaner environments for precision studies due to the absence of hadronic initial states and smaller experimental uncertainties. This structural simplicity is particularly beneficial when analysing processes such as top quark pair production, which yields a cleaner signal in hadronic final states. Lepton colliders are thus well-suited for precise measurements of the top quark mass and electroweak couplings via the $s$-channel pair production process. Proposed electron-positron machines like FCC-$ee$~\cite{FCC:2018byv,FCC:2018evy}, ILC~\cite{ILC:2013jhg}, CEPC~\cite{CEPCStudyGroup:2018rmc}, and CLIC~\cite{Lebrun:2012hj} hold significant potential for probing the top quark invariant mass lineshape should relevant centre-of-mass energies be achieved there, leveraging high luminosity and advanced detector capabilities. While the FCC-$ee$ excels near the $s$-channel $t\bar{t}$ production threshold due to its precise energy calibration, at higher energies, such as CLIC running at $\sqrt{s}=3~\text{TeV}$ becomes competitive, with its high luminosity potentially allowing a more detailed exploration of the lineshape over a wide range. However, significant beamstrahlung in CLIC can broaden the energy spectrum and reduce the precision in reconstructing the top invariant mass lineshape~\cite{Lebrun:2012hj,CLICdp:2018cto,Manosperti:2023sbd}. Muon colliders~\cite{AlAli:2021let}, on the other hand, offer distinct, exciting advantages over electron-positron colliders, as they can achieve higher centre-of-mass energies without significant synchrotron radiation losses, effectively making them vector boson colliders.  This enables muon colliders to probe top quark physics not only through $s$-channel pair production but also via weak boson fusion (WBF) processes, whose cross-section grows with centre-of-mass energy (see also~\cite{Chiesa:2020awd,Costantini:2020stv,Celada:2023oji}). This makes muon colliders ideal for capturing subtle variations in the top invariant mass lineshape influenced by the SMEFT operators discussed previously, hence potentially identifying new physics effects.
\subsubsection*{Event Simulation Details}
\begin{figure}[!t]
    \centering
    \parbox{5cm}{\begin{tikzpicture}
        \begin{feynman}
            \vertex (m1) at (-1, 1) {\(\large \mu^+\)};
            \vertex (m2) at (-1, -1) {\(\large \mu^-\)};
            \vertex (a) at (0, 0);
            \vertex (b) at (1, 0);
            \vertex (t1) at (2, 1);
            \vertex (t2) at (2, -1);
            \vertex (w1) at (3, 1.5) {\(\large W^+\)};
            \vertex (b1) at (3, 0.5) {\(\large b\)};
            \vertex (w2) at (3, -0.5) {\(\large W^-\)};
            \vertex (b2) at (3, -1.5) {\(\large \overline{b}\)};
            
            \diagram* {
                (m1) -- [anti fermion] (a) -- [anti fermion] (m2),
                (a) -- [boson, edge label'=\(Z/\gamma^*\)] (b),
                (b) -- [fermion, edge label=\(\large t\)] (t1),
                (b) -- [anti fermion, edge label=\(\large \overline{t}\)] (t2),
                (t1) -- [fermion] (b1),
                (t1) -- [boson] (w1),
                (t2) -- [anti fermion] (b2),
                (t2) -- [boson] (w2),
            };
        \end{feynman}
    \end{tikzpicture}}\hspace{1cm}
    \parbox{5.5cm}{ \begin{tikzpicture}
        \begin{feynman}
            \vertex (m1) at (-2.5, 1.5) {\(\large \mu^+\)};
            \vertex (m2) at (-2.5, -2.5) {\(\large \mu^-\)};
            \vertex (v1) at (-0.5,1);
            \vertex (a) at (0, 0);
            \vertex (b) at (0, -1);
            \vertex (v2) at (-0.5,-2);
            \vertex (t1) at (1, 0.25);
            \vertex (t2) at (1, -1.25);
            \vertex (w1) at (2, 1) {\(\large W^+\)};
            \vertex (b1) at (2, 0) {\(\large b\)};
            \vertex (w2) at (2, -1) {\(\large W^-\)};
            \vertex (b2) at (2, -2) {\(\large \overline{b}\)};
            \vertex (n1) at (1.5,1.5) {\(\large \overline{\nu}_{\mu}\)};
            \vertex (n2) at (1.5,-2.5) {\(\large \nu_{\mu}\)};
            
            \diagram* {
                (m1) -- [anti fermion] (v1),
                (v2) -- [anti fermion] (m2),
                (a) -- [fermion, edge label'=\(\large b\)] (b),
                (a) -- [fermion, edge label=\(\large t\)] (t1),
                (b) -- [anti fermion, edge label=\(\large \overline{t}\)] (t2),
                (a) -- [boson, edge label=\(\large W^+\)] (v1),
                (b) -- [boson, edge label'=\(\large W^-\)] (v2),
                (t1) -- [fermion] (b1),
                (t1) -- [boson] (w1),
                (t2) -- [anti fermion] (b2),
                (t2) -- [boson] (w2),
                (v1) -- [fermion] (n1),
                (v2) -- [fermion] (n2),
            };
        \end{feynman}
    \end{tikzpicture}}
    \caption{Representative Feynman topologies for $s$-channel and WBF $t \bar{t}$-production in $\mu^+ \mu^-$ colliders.\label{fig:ttbar}}
\end{figure}
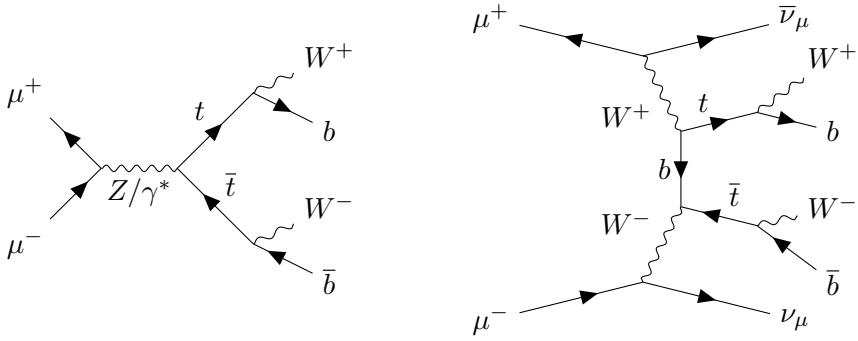
We use \texttt{SmeftFR}~\cite{Dedes:2017zog,Dedes:2019uzs,Dedes:2023zws} interfaced with \texttt{FeynRules}~\cite{Alloul:2013bka} to generate a \texttt{Ufo}~\cite{Degrande:2011ua} output containing the relevant Wilson coefficients, which we then link to \texttt{Madgraph5\_aMC@NLO}~\cite{Alwall:2011uj}. Events are generated by combining the processes $ \mu^+ \mu^- \rightarrow t \bar{t}$ and $ \mu^+ \mu^- \rightarrow t \bar{t} \nu_\mu \bar{\nu}_{\mu}$ to include both $s$-channel production of $t \bar{t}$ as well as top quark pair production through vector boson collisions. The top quark pairs are subsequently decayed semileptonically into light leptons ($e$, $\mu$), jets from the resulting $W$ bosons, and $b$ quarks. To incorporate the effects of a $q^2$-dependent running top quark width, we modify the top propagator as part of the \textsc{Helas} routines~\cite{Murayama:1992gi} entering the matrix element calculations for each value of the Wilson coefficients we examine (refer to App.~\ref{sec:helas} for implementation details).

Events are simulated for hypothetical future muon colliders with centre-of-mass energies of $\sqrt{s} = 3,10~\text{TeV}$. To reconstruct the top quark candidates, we closely follow the analyses performed in Ref.~\cite{Seidel:2013sqa,Chekanov:2017tfl,Li:2022iav} for semileptonically decaying $t \bar{t}$ candidates. We require exactly four jets for the final states, of which two are $b$-tagged, and exactly one light lepton with $p_T^l > 10~\text{GeV}$. To reconstruct the lineshape of the top quark candidate, we calculate the invariant mass of the two non $b$-tagged jets and the $b$-jet that is farther from the lepton. 
\subsubsection*{Bounds from Top Invariant Mass Lineshape}
We utilise the lineshape of the reconstructed top invariant mass distribution ($M_t$) to obtain a combined bound on the relevant SMEFT Wilson coefficients. The binned $\chi^2$ statistic for our analysis is constructed as
\begin{equation}
    \chi^2 (C_{tX}) =(b^i_{\text{SM}+\text{EFT}} (C_{tX}) - b^j_{\text{SM}}) V_{ij}^{-1} (b^j_{\text{SM}+\text{EFT}} (C_{tX}) - b^i_{\text{SM}}).
    \label{eqn:chi2}
\end{equation}
where $b^i_{\text{SM}+\text{EFT}} (C_{tX})$ denotes the combined number of events in the $i^{\text{th}}$ bin of the $M_t$ distribution, incorporating a running top quark width in the SMEFT framework and weighted by the cross-section, for a given value of $C_{tX}$, and, $b^i_{\text{SM}}$ corresponds to the number of events when considering only the SM running width. The covariance matrix $V_{ij}$ is constructed using the statistical Poisson uncertainty associated with each bin $b^{i}_{\text{SM}}$, along with a fully correlated relative fractional uncertainty ($\epsilon_r$). Thus, the covariance matrix can be expressed as
\begin{equation}
    V_{ij} = b^i_{\text{SM}} \delta_{ij} + \epsilon_r b^i_{\text{SM}} b^j_{\text{SM}}.
    \label{eqn:covmat}
\end{equation} 
The $95\%$ confidence level bounds on $C_{tX} = C_{tW} = C_{tB} = C_{tG} = C_{t\phi}$ were calculated for $\epsilon_r = 10\%$ and $25\%$. Our results for different centre-of-mass energies and integrated luminosities are presented in Tab.~\ref{tab:minvbds}. 

How do these bounds compare to global fit analyses? The operators $C_{tG}$, and $C_{t\phi}$ are well constrained in hadron collider processes~\cite{Celada:2024mcf}; however, $C_{tW}$ and $C_{tB}$ can be directly probed at lepton colliders near the $t\bar{t}$-production threshold. We adopt the methodologies employed in Refs.~\cite{Durieux:2018tev,CLICdp:2018esa} and \textsc{TopFitter}~\cite{Buckley:2015lku} for $e^+ e^-$ colliders (see also~\cite{Englert:2017dev}), adapting them for muon colliders to assess whether the top invariant mass lineshape can effectively constrain these operators more robustly than global fit analyses.

\begin{table}[!t]
    \centering
    \begin{tabular}{|c||c|c|c|}
    \hline
    $\mathcal{L}$ & $\epsilon_r$ & $\sqrt{s}=3~\text{TeV}$ & $\sqrt{s}=10~\text{TeV}$ \\ \hline \hline
    \multirow{2}{4em}{$5~\text{ab}^{-1}$} & 25\% & $(-0.83,1.12)~\text{TeV}^{-2}$ & $(-0.52,0.48)~\text{TeV}^{-2}$ \\
    & 10\% & $(-0.54,0.60)~\text{TeV}^{-2}$ & $(-0.23,0.21)~\text{TeV}^{-2}$ \\ \hline
    \multirow{2}{4em}{$10~\text{ab}^{-1}$} & 25\% & $(-0.72,0.91)~\text{TeV}^{-2}$ & $(-0.42,0.41)~\text{TeV}^{-2}$ \\ 
    & 10\% & $(-0.51,0.57)~\text{TeV}^{-2}$ & $(-0.20,0.18)~\text{TeV}^{-2}$ \\ \hline
    \end{tabular}
    \caption{95\% CL bounds on the relevant SMEFT Wilson coefficients ($C_{tX}$) from the top invariant mass lineshape at a future muon collider.}
    \label{tab:minvbds}
\end{table}

Firstly, we evaluate the sensitivity of both the total cross-section ($\sigma$) and the forward-backward asymmetry ($A_{\text{FB}}$) to the effective operators influencing $t \bar{t}$ production at muon colliders. The forward-backward asymmetry is defined as
\begin{equation}
    A_{\text{FB}} = \frac{\sigma_{\text{FB}}}{\sigma},
\end{equation}
where the forward-backward cross-section $\sigma_{\text{FB}}$ is given by
\begin{equation}
    \sigma_{\text{FB}} = \int^{1}_{-1} d \cos{\theta_t}~\text{sign}(\cos{\theta_t})~\frac{d \sigma}{d \cos{\theta_t}}
\end{equation}
with $\theta_t$ representing the scattering angle in the centre-of-mass frame. Events are generated in \texttt{Madgraph5\_aMC@NLO} for various initial beam polarisations of the $\mu^+ \mu^-$ states. From the resulting events, we compute the SM, linear, and quadratic dependences of the effective operators on the respective observables.

\begin{figure}[!t]
    \centering
    \includegraphics[width=0.49\linewidth]{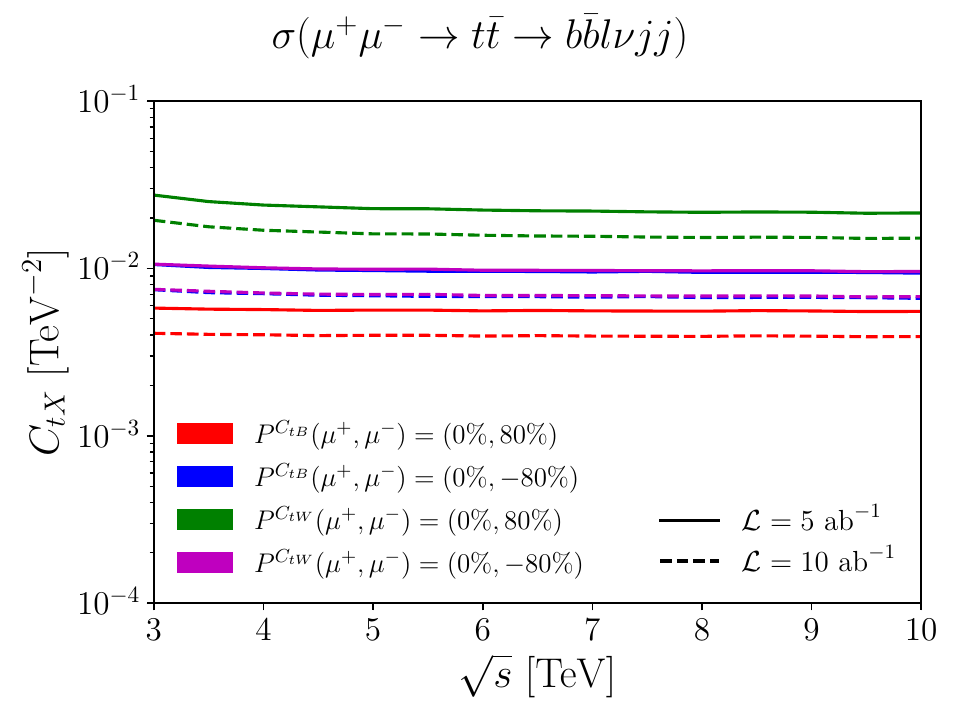}
    \includegraphics[width=0.49\linewidth]{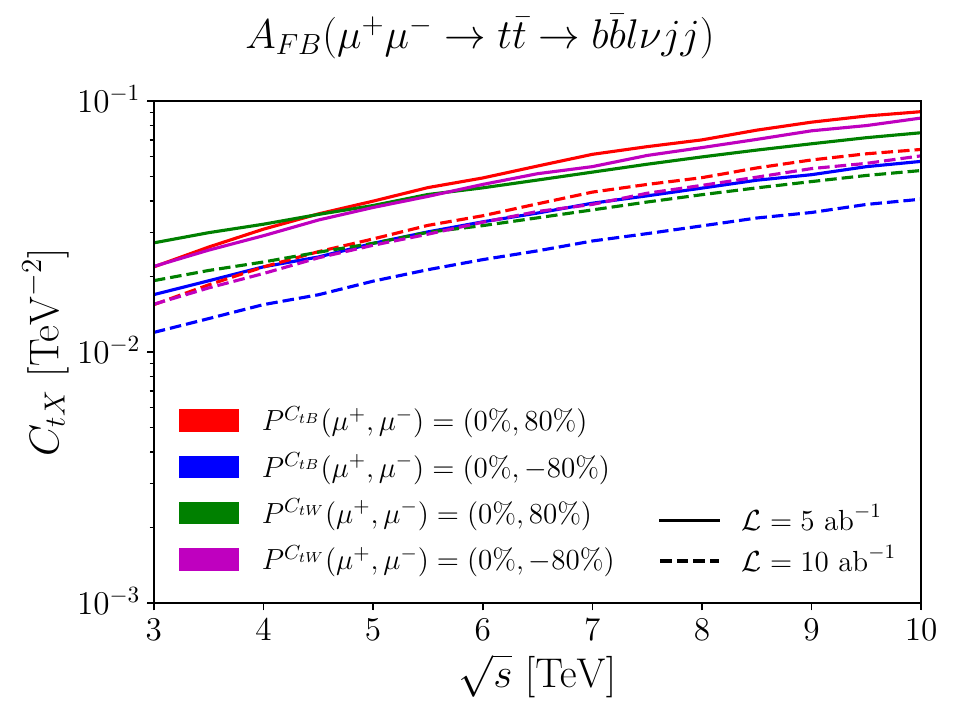}
    \caption{95\% CL upper bounds on $C_{tW}$ and $C_{tB}$ from (left) cross-section and (right) forward-backward asymmetry~\cite{Durieux:2018tev,CLICdp:2018esa} for different initial beam polarisations. \label{fig:gfitbds}}
\end{figure}

The sensitivity of an observable $o$ to a Wilson coefficient $C_i$ ($S^o_i$) is calculated as per Ref.~\cite{Durieux:2018tev}, as its normalised variation around the SM point:
\begin{equation}
    S_i^o = \left.\frac{1}{o}\frac{\partial o}{\partial C_i}\right|_{C_i = 0} = \frac{o_i}{o_{\text{SM}}},\quad\text{with}\quad o = o_{\text{SM}} + C_i o_i + C_i C_j o_{ij} + ... \quad .
\end{equation}
Here, we have conveniently set the cutoff scale $\Lambda = 1~\text{TeV}$, which can also be absorbed into the definitions of $o_i$, $o_{ij}$, etc. We then derive 95\% confidence limit constraints on $C_{tB}$ and $C_{tW}$ from the sensitivities of the cross-section and forward-backward asymmetry as functions of centre-of-mass energy for different initial beam polarisations and integrated luminosities. Figure~\ref{fig:gfitbds} shows that these constraints offer substantially stronger bounds on the respective effective operators compared to those obtained from the top invariant mass lineshape.
We also apply the methodology of \textsc{TopFitter}~\cite{Buckley:2015lku} to cross-check the abovementioned bounds. Following Ref.~\cite{Buckley:2015lku}, and motivated by the dependence of the total cross-section of the process $\mu^+ \mu^- \rightarrow t \bar{t}$ on the Wilson coefficients ($\sigma[C_{tX}]$), we fit a polynomial of degree greater than two. Without systematic uncertainties, each observable would ideally follow a second-order polynomial in the coefficients; higher-order terms account for bin uncertainties that introduce deviations from this ideal form. To determine constraints on the Wilson coefficients, we construct a $\chi^2$ distribution as
\begin{equation}
    \chi^2 (C_{tX}) = \frac{(\sigma[C_{tX}] - \sigma)^2}{\sigma},
\end{equation}
where $\sigma = \sigma_{\text{SM}} (1 + \Sigma)$. Here, $\sigma_{\text{SM}}$ represents the SM cross-section for the same process and $\Sigma$ represents the combined theoretical and experimental uncertainties associated with the cross-section. Assuming a 10\% combined uncertainty in the SM cross-section, we obtain 95\% confidence limit constraints for $C_{tW}$ and $C_{tB}$ at various centre-of-mass energies $\sqrt{s}$ and integrated luminosities, as presented in Tab.~\ref{tab:gfitbds}. These bounds further demonstrate that direct constraints from global fit analyses are more effective in constraining the relevant Wilson coefficients than those derived from the top invariant mass lineshape.
\begin{table}[!t]
    \centering
    \begin{tabular}{|c||c|c|c|}
    \hline
    $\mathcal{L}$ & Wilson coefficient & $\sqrt{s}=3~\text{TeV}$ & $\sqrt{s}=10~\text{TeV}$ \\ \hline \hline
    \multirow{2}{4em}{$5~\text{ab}^{-1}$} & $C_{tB}$ & $(-0.02,0.15)~\text{TeV}^{-2}$ & $(-0.02,0.04)~\text{TeV}^{-2}$ \\
    & $C_{tW}$ & $(-0.02,0.24)~\text{TeV}^{-2}$ & $(-0.03,0.05)~\text{TeV}^{-2}$ \\ \hline
    \multirow{2}{4em}{$10~\text{ab}^{-1}$} & $C_{tB}$ & $(-0.02,0.14)~\text{TeV}^{-2}$ & $(-0.02,0.03)~\text{TeV}^{-2}$ \\
    & $C_{tW}$ & $(-0.02,0.23)~\text{TeV}^{-2}$ & $(-0.02,0.05)~\text{TeV}^{-2}$ \\ \hline
    \end{tabular}
    \caption{95\% CL individual bounds on the relevant SMEFT Wilson coefficients from the \textsc{TopFitter}~\cite{Buckley:2015lku} analysis.}
    \label{tab:gfitbds}
\end{table}

\section{Conclusions}
\label{sec:conclusion}
Width pseudo-observables are hallmarks for past, present and future particle collider experiments. Peak values of cross sections and branching ratios are essential to verify the Standard Model hypothesis further, measure SM particle properties, and, most importantly, inform the search for new physics beyond the SM by revealing deviations from precise SM theory predictions. Momentum dependencies of widths in the SM are often not considered in full in such investigations, and for a good reason, as these play a subdominant role given the past and current experimental sensitivity. With future precision machines under increasing discussion and consideration, the extent to which such nuances are missed is less clear. In parallel, informed by the sensitivity expectations for future hadron or precision lepton machines, the impact of heavy physics reshaping the SM's behaviour in this context is poorly understood. Our investigation aims to close this gap and provide a quantitative estimate of the relevance of lineshape deviations in the light of expected direct sensitivity. To this end, we have parametrised new heavy physics by means of SMEFT, focussing on electroweak-scale particles of the SM, such as the $W$ boson, the Higgs boson, and the top quark.

For bosonic particles, e.g. the Higgs boson or the $W$ boson, the running width in the SM deviates slightly from the fixed-width approximation, with percent-level effects visible in the reconstructed transverse mass distribution, see Fig.~\ref{fig:mtw}. The momentum-dependent $W$ width introduces corrections to the invariant mass lineshape, which, while subtle, could become relevant for precision measurements of the $W$ boson mass, such as those reported by the ATLAS and CMS collaborations. For the Higgs boson, due to its much smaller intrinsic width relative to its mass, the impact of momentum dependence is negligible, as evident from the close agreement between the fixed and running propagators shown in Fig.~\ref{fig:hplot}. These findings suggest that momentum dependence is not critical for Higgs boson analyses under current experimental sensitivities. However, the $W$ boson could serve as a more sensitive probe for such effects in the precision era. As new physics needs to be considered a perturbation within the theoretical boundaries of our analysis, the relevance to BSM physics of the Higgs and $W$ boson width momentum dependencies is not directly clear; measurements of the related pseudo-observables will likely be more sensitive as they currently are already pursued.

The top quark analysis revealed significantly larger effects due to its relatively broad width and dynamic self-energy contributions. We observed pronounced deviations between the fixed-width and running-width propagators, particularly above the top quark mass threshold, as shown in Fig.~\ref{fig:tplot}. These deviations arise from the energy dependence of the imaginary parts of the self-energy components (i.e., $\Sigma_L$, $\Sigma_R$, and $\Sigma_S$), which exhibit notable changes near $q^2 = m_t^2$. These effects become significantly pronounced when SMEFT operators are introduced, as they modify the self-energy contributions and, consequently, the top quark's invariant mass lineshape, as illustrated in Fig.~\ref{fig:tploteft}. This highlights the top quark yet again as an outstanding candidate for a new physics discovery if its SM hypothesis can be brought under sufficient theoretical control.

Therefore, we further explored this in the context of collider experiments. At hadron colliders, such as the LHC, the reconstructed invariant mass distributions for the top quark exhibited limited sensitivity to running-width effects, with percent-level deviations observed in Fig.~\ref{fig:mlb}. This is primarily due to the challenging experimental environment, where detector effects and uncertainties dominate. However, future lepton colliders, such as proposed muon colliders operating at centre-of-mass energies of 3 and 10 TeV, present a much cleaner environment for such studies. Our simulations demonstrated that lepton colliders could achieve significantly improved constraints on SMEFT-induced deviations in the top propagator and running width. Our constraints pass quality checks related to perturbative unitarity (see Appendix~\ref{sec:unitarity}). The bounds obtained through the impact of the top quark's running width on the invariant mass lineshape can complement those derived from global fit analyses (Tab.~\ref{tab:gfitbds}), providing another avenue to probing new physics. The observed deviations in the running width at high $q^2$, particularly when including operators like $O_{tG}$ and $O_{tW}$, emphasise the potential of collider experiments to detect subtle effects beyond the SM.

Our study demonstrates the relevance of momentum-dependent widths in theoretical predictions, particularly as the field moves toward higher precision in both measurements and theoretical calculations. While some effects remain below current experimental sensitivities, the next generation of lepton colliders will likely probe these phenomena with unprecedented precision. This will refine our understanding of SM particle properties and enhance the discovery potential for physics beyond the Standard Model.

\subsection*{Acknowledgements}
We thank Dave Sutherland for helpful and insightful discussions. C.E. and M.S. thank the DESY theory and CMS experimental groups, particularly G.~Weiglein, P.~Stylianou, and A.~Belvedere, R.~Kogler, C.~Schwanenberger, for their hospitality during a visit to Hamburg, where part of this work was completed. C.E. is supported by the STFC under grant ST/X000605/1 and by the Leverhulme Trust under Research Project Grant RPG-2021-031 and Research Fellowship RF-2024-300$\backslash$9. C.E. is further supported by the Institute for Particle Physics Phenomenology Associateship Scheme. W.N. is funded by a University of Glasgow College of Science and Engineering Scholarship. M.S. is supported by the STFC under grant ST/P001246/1.
\appendix
\section{Perturbative Unitarity bounds on the SMEFT Wilson coefficients}
\label{sec:unitarity}
Unitarity provides a suitable tool to gauge whether bounds obtained on the SMEFT Wilson coefficients are within the perturbative regime. We consider partial wave unitarity constraints from $\gamma t \rightarrow \gamma t$ scattering to calculate the unitarity bounds on the Wilson coefficients, considering $(\text{dim-6})^2$ insertions. The zeroth partial wave for $i_1 i_2 \rightarrow f_1 f_2$ scattering is given by
\begin{equation}
    a_{fi}^0 =\frac{1}{32 \pi s}\beta^{\frac{1}{4}}\left(s,m^2_{i_1},m^2_{i_2}\right)\beta^{\frac{1}{4}}\left(s,m^2_{f_1},m^2_{f_2}\right) \int_{-1}^{1} d\cos{\theta}\, \mathcal{M}(\sqrt{s},\cos{\theta}),
\end{equation}
where $\sqrt{s}$ represents the centre-of-mass energy, $\theta$ is the scattering angle for the $2 \rightarrow 2$ process described by $\mathcal{M}$ in the centre-of-mass frame, and $\beta$ is defined as
\begin{equation*}
    \beta (x,y,z) = x^2 + y^2 + z^2 - 2 x y - 2 y z - 2 z x.
\end{equation*} 
\begin{figure}[!t]
    \centering
    \includegraphics[width=0.6\linewidth]{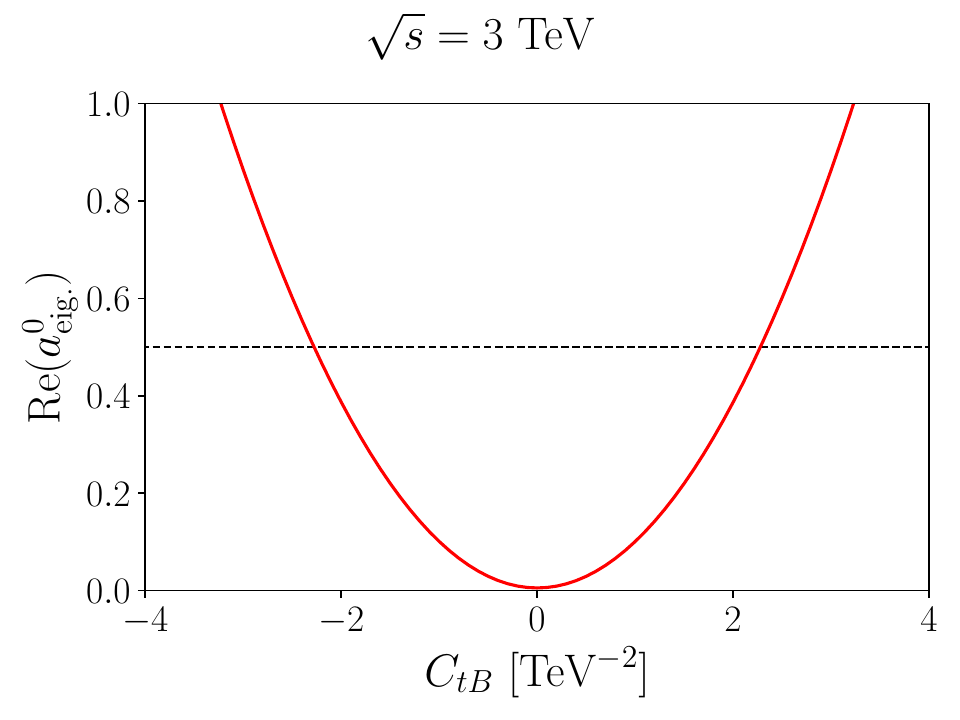}
    \caption{Perturbative unitarity bounds on $C_{tB}$.\label{fig:unitarity}}
\end{figure}
In the case of $\gamma t \rightarrow \gamma t$, we construct a matrix of zeroth partial waves, restricting ourselves to states with helicity equal to zero. The unitarity condition on the $S$-matrix then translates to
\begin{equation}
    |\text{Re} (a^0_{\text{eig.}})| < \frac{1}{2},\quad\text{and}\quad |\text{Im} (a^0_{\text{eig.}})| < 1.
\end{equation}
where $a^0_{\text{eig.}}$ is the largest eigenvalue of the matrix constructed from the zeroth partial waves. We use the former condition to obtain the unitarity bounds. Figure~\ref{fig:unitarity} shows $\text{Re} (a^0_{\text{eig.}})$ as a function of the Wilson coefficient $C_{tB}$ for $\sqrt{s} = 3~\text{TeV}$, from which we infer that the constraints obtained in previous sections are indeed within the perturbative regime.
\section{Implementation of the running top quark width on \textsc{HELAS} routines}
\label{sec:helas}
The \textsc{Helas} (HELicity Amplitude Subroutines) routines~\cite{Murayama:1992gi} are responsible for generating matrix elements from Feynman rules in Monte-Carlo event generators like \textsc{MadGraph5}\_\textsc{aMC}@\textsc{NLO}. By modifying the {\textsc{Helas}} routines directly, their gauge invariance properties (i.e. unitary gauge) at the considered LO approximation extend to our modifications. In the standard \textsc{Helas} implementation, the top quark propagator follows the fixed-width Breit-Wigner form 
\begin{equation}
    i G^t(\pslash,q^2) = i\frac{\pslash + m_t}{q^2 - m_t^2 + i m_t \Gamma_t}.
\end{equation}
As discussed in Sec.~\ref{sec:top}, in the narrow-width approximation, the momentum-dependent width affects only the denominator of the full propagator. The relevant propagator-denominator information in the subroutines is encoded in the \texttt{DENOM} variable. For example, in the fixed-width case, the implementation in a subroutine appears as:
\begin{lstlisting}[columns=fullflexible]
  DENOM = COUP/(P1(0)**2-P1(1)**2-P1(2)**2-P1(3)**2 - M1 * (M1 - CI * W1)).
\end{lstlisting}
Here, the fixed Breit-Wigner width is encoded in the ``\texttt{W1}" variable. To incorporate the dynamic momentum-dependent width across all top quark propagators, we followed a systematic approach:
\begin{enumerate}
    \item We searched for subroutines where the top quark width is explicitly passed as an argument. Since we focus on the $t \rightarrow W b$ decay, the relevant subroutines belong to the class ``\texttt{FFVXXX.f}", which governs fermion-fermion-gauge boson interactions.
    \item We replaced the fixed-width parameter (for example, \texttt{W1}) with the $q^2$-dependent width, which can be computed either directly from the analytical expression in Eq.~(\ref{eq:wid3}) or via polynomial interpolation as a function of $q^2$.
    \item Additional conditional statements were introduced to correctly handle cases where the propagator appears in the $t$-channel or when the same subroutine is used for different particles (e.g., the $b$-quark).
\end{enumerate}
This implementation ensures that the running width is consistently applied in all relevant interactions in a numerically stable way. The same approach can be applied to incorporate momentum-dependent running widths in the presence of SMEFT insertions.
\begin{figure}[!t]
    \centering
    \subfigure[\label{subfig:ppcol} Hadron collider.]{\includegraphics[height=0.35\linewidth]{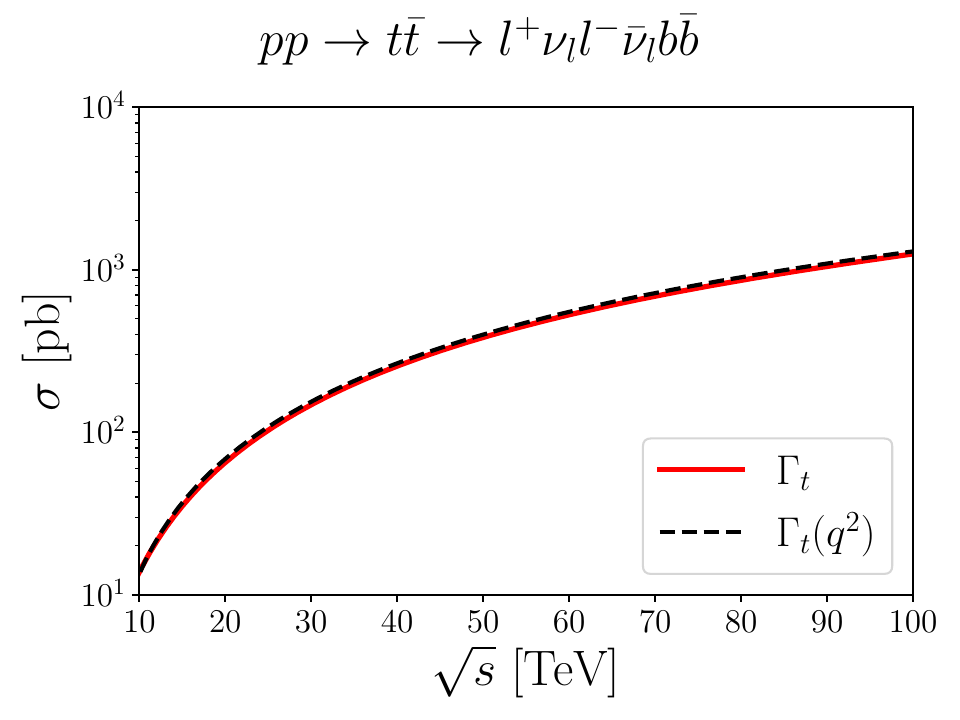}}\hfill
    \subfigure[\label{subfig:mmcol} Muon collider.]{\includegraphics[height=0.35\linewidth]{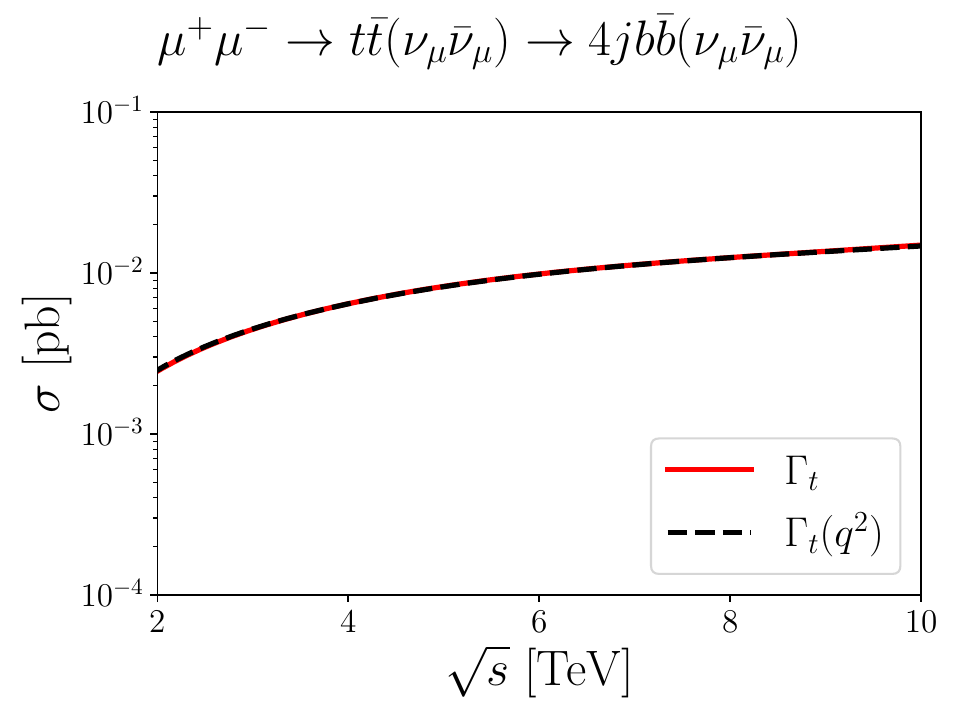}}
    \caption{Plots showing the cross-section dependence of top quark pair production on the fixed and momentum-dependent top quark width widths across the different centre-of-mass energies at a \textbf{(a)} $p p $ hadron collider and a \textbf{(b)} muon collider. The muon collider process includes both the $s$-channel and WBF $t \bar{t}$ production modes. \label{fig:colplot}}
\end{figure}
To validate our implementation, we cross-checked the computed cross-sections using the SM fixed and running widths for the relevant processes at both hadron and muon colliders, ensuring that no unphysical behavior emerged. We verified that the inclusion of the momentum-dependent width does not introduce spurious effects in the total cross-section calculations. The results, shown in Fig.~\ref{fig:colplot}, demonstrate consistency across both $pp$ and muon colliders (including the WBF channel). Furthermore, we observe no unitarity-violating growth in cross-sections at high centre-of-mass energies, confirming the stability of our implementation. 
\bibliographystyle{JHEP}
\bibliography{references}

\providecommand{\href}[2]{#2}\begingroup\raggedright\begin{thebibliography}{10}

\bibitem{Nowakowski:1993iu}
M.~Nowakowski and A.~Pilaftsis, \emph{{On gauge invariance of Breit-Wigner
  propagators}}, \href{https://doi.org/10.1007/BF01650437}{\emph{Z. Phys. C}
  {\bfseries 60} (1993) 121}
  [\href{https://arxiv.org/abs/hep-ph/9305321}{{\ttfamily hep-ph/9305321}}].

\bibitem{Denner:2019vbn}
A.~Denner and S.~Dittmaier, \emph{{Electroweak Radiative Corrections for
  Collider Physics}},
  \href{https://doi.org/10.1016/j.physrep.2020.04.001}{\emph{Phys. Rept.}
  {\bfseries 864} (2020) 1} [\href{https://arxiv.org/abs/1912.06823}{{\ttfamily
  1912.06823}}].

\bibitem{Lehmann:1954rq}
H.~Lehmann, K.~Symanzik and W.~Zimmermann, \emph{{On the formulation of
  quantized field theories}},
  \href{https://doi.org/10.1007/BF02731765}{\emph{Nuovo Cim.} {\bfseries 1}
  (1955) 205}.

\bibitem{1976opth}
R.~G. Newton, \emph{{Optical theorem and beyond}},
  \href{https://doi.org/10.1119/1.10324}{\emph{American Journal of Physics}
  (1976) }.

\bibitem{Cutkosky:1960sp}
R.~E. Cutkosky, \emph{{Singularities and discontinuities of Feynman
  amplitudes}}, \href{https://doi.org/10.1063/1.1703676}{\emph{J. Math. Phys.}
  {\bfseries 1} (1960) 429}.

\bibitem{Veltman:1963th}
M.~J.~G. Veltman, \emph{{Unitarity and causality in a renormalizable field
  theory with unstable particles}},
  \href{https://doi.org/10.1016/S0031-8914(63)80277-3}{\emph{Physica}
  {\bfseries 29} (1963) 186}.

\bibitem{Passarino:2010qk}
G.~Passarino, C.~Sturm and S.~Uccirati, \emph{{Higgs Pseudo-Observables, Second
  Riemann Sheet and All That}},
  \href{https://doi.org/10.1016/j.nuclphysb.2010.03.013}{\emph{Nucl. Phys. B}
  {\bfseries 834} (2010) 77} [\href{https://arxiv.org/abs/1001.3360}{{\ttfamily
  1001.3360}}].

\bibitem{Weldon:1975gu}
H.~A. Weldon, \emph{{The Description of Unstable Particles in Quantum Field
  Theory}}, \href{https://doi.org/10.1103/PhysRevD.14.2030}{\emph{Phys. Rev. D}
  {\bfseries 14} (1976) 2030}.

\bibitem{Stuart:1991xk}
R.~G. Stuart, \emph{{Gauge invariance, analyticity and physical observables at
  the Z0 resonance}},
  \href{https://doi.org/10.1016/0370-2693(91)90653-8}{\emph{Phys. Lett. B}
  {\bfseries 262} (1991) 113}.

\bibitem{Aeppli:1993rs}
A.~Aeppli, G.~J. van Oldenborgh and D.~Wyler, \emph{{Unstable particles in one
  loop calculations}},
  \href{https://doi.org/10.1016/0550-3213(94)90195-3}{\emph{Nucl. Phys. B}
  {\bfseries 428} (1994) 126}
  [\href{https://arxiv.org/abs/hep-ph/9312212}{{\ttfamily hep-ph/9312212}}].

\bibitem{Seymour:1995np}
M.~H. Seymour, \emph{{The Higgs boson line shape and perturbative unitarity}},
  \href{https://doi.org/10.1016/0370-2693(95)00699-L}{\emph{Phys. Lett. B}
  {\bfseries 354} (1995) 409}
  [\href{https://arxiv.org/abs/hep-ph/9505211}{{\ttfamily hep-ph/9505211}}].

\bibitem{Baur:1995aa}
U.~Baur and D.~Zeppenfeld, \emph{{Finite width effects and gauge invariance in
  radiative $W$ productions and decay}},
  \href{https://doi.org/10.1103/PhysRevLett.75.1002}{\emph{Phys. Rev. Lett.}
  {\bfseries 75} (1995) 1002}
  [\href{https://arxiv.org/abs/hep-ph/9503344}{{\ttfamily hep-ph/9503344}}].

\bibitem{Nielsen:1975fs}
N.~K. Nielsen, \emph{{On the Gauge Dependence of Spontaneous Symmetry Breaking
  in Gauge Theories}},
  \href{https://doi.org/10.1016/0550-3213(75)90301-6}{\emph{Nucl. Phys. B}
  {\bfseries 101} (1975) 173}.

\bibitem{Gambino:1999ai}
P.~Gambino and P.~A. Grassi, \emph{{The Nielsen identities of the SM and the
  definition of mass}},
  \href{https://doi.org/10.1103/PhysRevD.62.076002}{\emph{Phys. Rev. D}
  {\bfseries 62} (2000) 076002}
  [\href{https://arxiv.org/abs/hep-ph/9907254}{{\ttfamily hep-ph/9907254}}].

\bibitem{Grassi:2001bz}
P.~A. Grassi, B.~A. Kniehl and A.~Sirlin, \emph{{Width and partial widths of
  unstable particles in the light of the Nielsen identities}},
  \href{https://doi.org/10.1103/PhysRevD.65.085001}{\emph{Phys. Rev. D}
  {\bfseries 65} (2002) 085001}
  [\href{https://arxiv.org/abs/hep-ph/0109228}{{\ttfamily hep-ph/0109228}}].

\bibitem{Goria:2011wa}
S.~Goria, G.~Passarino and D.~Rosco, \emph{{The Higgs Boson Lineshape}},
  \href{https://doi.org/10.1016/j.nuclphysb.2012.07.006}{\emph{Nucl. Phys. B}
  {\bfseries 864} (2012) 530}
  [\href{https://arxiv.org/abs/1112.5517}{{\ttfamily 1112.5517}}].

\bibitem{Sirlin:1991fd}
A.~Sirlin, \emph{{Theoretical considerations concerning the $Z^0$ mass}},
  \href{https://doi.org/10.1103/PhysRevLett.67.2127}{\emph{Phys. Rev. Lett.}
  {\bfseries 67} (1991) 2127}.

\bibitem{Denner:1999gp}
A.~Denner, S.~Dittmaier, M.~Roth and D.~Wackeroth, \emph{{Predictions for all
  processes $e^+ e^- \to 4\;\text{fermions} + \gamma$}},
  \href{https://doi.org/10.1016/S0550-3213(99)00437-X}{\emph{Nucl. Phys. B}
  {\bfseries 560} (1999) 33}
  [\href{https://arxiv.org/abs/hep-ph/9904472}{{\ttfamily hep-ph/9904472}}].

\bibitem{Denner:2006ic}
A.~Denner and S.~Dittmaier, \emph{{The complex-mass scheme for perturbative
  calculations with unstable particles}},
  \href{https://doi.org/10.1016/j.nuclphysbps.2006.09.025}{\emph{Nucl. Phys. B
  Proc. Suppl.} {\bfseries 160} (2006) 22}
  [\href{https://arxiv.org/abs/hep-ph/0605312}{{\ttfamily hep-ph/0605312}}].

\bibitem{Actis:2006rc}
S.~Actis and G.~Passarino, \emph{{Two-Loop Renormalization in the Standard
  Model Part III: Renormalization Equations and their Solutions}},
  \href{https://doi.org/10.1016/j.nuclphysb.2007.04.027}{\emph{Nucl. Phys. B}
  {\bfseries 777} (2007) 100}
  [\href{https://arxiv.org/abs/hep-ph/0612124}{{\ttfamily hep-ph/0612124}}].

\bibitem{Wetzel:1982mh}
W.~Wetzel, \emph{{Electroweak Radiative Corrections for $e^+ e^- \to \mu^+
  \mu^-$ at LEP Energies}},
  \href{https://doi.org/10.1016/0550-3213(83)90139-6}{\emph{Nucl. Phys. B}
  {\bfseries 227} (1983) 1}.

\bibitem{Berends:1987ab}
F.~A. Berends, W.~L. van Neerven and G.~J.~H. Burgers, \emph{{Higher Order
  Radiative Corrections at LEP Energies}},
  \href{https://doi.org/10.1016/0550-3213(88)90313-6}{\emph{Nucl. Phys. B}
  {\bfseries 297} (1988) 429}.

\bibitem{Beenakker:1988pv}
W.~Beenakker and W.~Hollik, \emph{{The Width of the Z Boson}},
  \href{https://doi.org/10.1007/BF01559728}{\emph{Z. Phys. C} {\bfseries 40}
  (1988) 141}.

\bibitem{Bardin:1988xt}
D.~Y. Bardin, A.~Leike, T.~Riemann and M.~Sachwitz, \emph{{Energy Dependent
  Width Effects in $e^+ e^-$ Annihilation Near the $Z$ Boson Pole}},
  \href{https://doi.org/10.1016/0370-2693(88)91627-9}{\emph{Phys. Lett. B}
  {\bfseries 206} (1988) 539}.

\bibitem{Papavassiliou:1995fq}
J.~Papavassiliou and A.~Pilaftsis, \emph{{Gauge invariance and unstable
  particles}}, \href{https://doi.org/10.1103/PhysRevLett.75.3060}{\emph{Phys.
  Rev. Lett.} {\bfseries 75} (1995) 3060}
  [\href{https://arxiv.org/abs/hep-ph/9506417}{{\ttfamily hep-ph/9506417}}].

\bibitem{Passera:1996nk}
M.~Passera and A.~Sirlin, \emph{{Analysis of the $Z^0$ resonant amplitude in
  the general $R(\xi)$ gauges}},
  \href{https://doi.org/10.1103/PhysRevLett.77.4146}{\emph{Phys. Rev. Lett.}
  {\bfseries 77} (1996) 4146}
  [\href{https://arxiv.org/abs/hep-ph/9607253}{{\ttfamily hep-ph/9607253}}].

\bibitem{Denner:2014zga}
A.~Denner and J.-N. Lang, \emph{{The Complex-Mass Scheme and Unitarity in
  perturbative Quantum Field Theory}},
  \href{https://doi.org/10.1140/epjc/s10052-015-3579-2}{\emph{Eur. Phys. J. C}
  {\bfseries 75} (2015) 377} [\href{https://arxiv.org/abs/1406.6280}{{\ttfamily
  1406.6280}}].

\bibitem{Maltoni:2018zvp}
F.~Maltoni, M.~K. Mandal and X.~Zhao, \emph{{Top-quark effects in diphoton
  production through gluon fusion at next-to-leading order in QCD}},
  \href{https://doi.org/10.1103/PhysRevD.100.071501}{\emph{Phys. Rev. D}
  {\bfseries 100} (2019) 071501}
  [\href{https://arxiv.org/abs/1812.08703}{{\ttfamily 1812.08703}}].

\bibitem{Smith:1996xz}
M.~C. Smith and S.~S. Willenbrock, \emph{{Top quark pole mass}},
  \href{https://doi.org/10.1103/PhysRevLett.79.3825}{\emph{Phys. Rev. Lett.}
  {\bfseries 79} (1997) 3825}
  [\href{https://arxiv.org/abs/hep-ph/9612329}{{\ttfamily hep-ph/9612329}}].

\bibitem{Beneke:1994sw}
M.~Beneke and V.~M. Braun, \emph{{Heavy quark effective theory beyond
  perturbation theory: Renormalons, the pole mass and the residual mass term}},
  \href{https://doi.org/10.1016/0550-3213(94)90314-X}{\emph{Nucl. Phys. B}
  {\bfseries 426} (1994) 301}
  [\href{https://arxiv.org/abs/hep-ph/9402364}{{\ttfamily hep-ph/9402364}}].

\bibitem{Bigi:1994em}
I.~I.~Y. Bigi, M.~A. Shifman, N.~G. Uraltsev and A.~I. Vainshtein, \emph{{The
  Pole mass of the heavy quark. Perturbation theory and beyond}},
  \href{https://doi.org/10.1103/PhysRevD.50.2234}{\emph{Phys. Rev. D}
  {\bfseries 50} (1994) 2234}
  [\href{https://arxiv.org/abs/hep-ph/9402360}{{\ttfamily hep-ph/9402360}}].

\bibitem{Dittmaier:2009cr}
S.~Dittmaier and M.~Huber, \emph{{Radiative corrections to the neutral-current
  Drell-Yan process in the Standard Model and its minimal supersymmetric
  extension}}, \href{https://doi.org/10.1007/JHEP01(2010)060}{\emph{JHEP}
  {\bfseries 01} (2010) 060} [\href{https://arxiv.org/abs/0911.2329}{{\ttfamily
  0911.2329}}].

\bibitem{Dittmaier:2014qza}
S.~Dittmaier, A.~Huss and C.~Schwinn, \emph{{Mixed QCD-electroweak
  $\mathcal{O}(\alpha_s\alpha)$ corrections to Drell-Yan processes in the
  resonance region: pole approximation and non-factorizable corrections}},
  \href{https://doi.org/10.1016/j.nuclphysb.2014.05.027}{\emph{Nucl. Phys. B}
  {\bfseries 885} (2014) 318}
  [\href{https://arxiv.org/abs/1403.3216}{{\ttfamily 1403.3216}}].

\bibitem{Kauer:2002sn}
N.~Kauer, \emph{{Top pair production beyond double pole approximation: $p p,p
  \bar{p} \to$ six fermions and zero, one or two additional partons}},
  \href{https://doi.org/10.1103/PhysRevD.67.054013}{\emph{Phys. Rev. D}
  {\bfseries 67} (2003) 054013}
  [\href{https://arxiv.org/abs/hep-ph/0212091}{{\ttfamily hep-ph/0212091}}].

\bibitem{Denner:1991kt}
A.~Denner, \emph{{Techniques for calculation of electroweak radiative
  corrections at the one loop level and results for W physics at LEP-200}},
  \href{https://doi.org/10.1002/prop.2190410402}{\emph{Fortsch. Phys.}
  {\bfseries 41} (1993) 307} [\href{https://arxiv.org/abs/0709.1075}{{\ttfamily
  0709.1075}}].

\bibitem{ParticleDataGroup:2024cfk}
{\scshape Particle Data Group} collaboration, \emph{{Review of particle
  physics}}, \href{https://doi.org/10.1103/PhysRevD.110.030001}{\emph{Phys.
  Rev. D} {\bfseries 110} (2024) 030001}.

\bibitem{Englert:2015zra}
C.~Englert, I.~Low and M.~Spannowsky, \emph{{On-shell interference effects in
  Higgs boson final states}},
  \href{https://doi.org/10.1103/PhysRevD.91.074029}{\emph{Phys. Rev. D}
  {\bfseries 91} (2015) 074029}
  [\href{https://arxiv.org/abs/1502.04678}{{\ttfamily 1502.04678}}].

\bibitem{Kauer:2015hia}
N.~Kauer and C.~O'Brien, \emph{{Heavy Higgs signal\textendash{}background
  interference in $gg\rightarrow VV$ in the Standard Model plus real singlet}},
  \href{https://doi.org/10.1140/epjc/s10052-015-3586-3}{\emph{Eur. Phys. J. C}
  {\bfseries 75} (2015) 374}
  [\href{https://arxiv.org/abs/1502.04113}{{\ttfamily 1502.04113}}].

\bibitem{deBlas:2019rxi}
J.~de~Blas et~al., \emph{{Higgs Boson Studies at Future Particle Colliders}},
  \href{https://doi.org/10.1007/JHEP01(2020)139}{\emph{JHEP} {\bfseries 01}
  (2020) 139} [\href{https://arxiv.org/abs/1905.03764}{{\ttfamily
  1905.03764}}].

\bibitem{Dawson:2022zbb}
S.~Dawson et~al., \emph{{Report of the Topical Group on Higgs Physics for
  Snowmass 2021: The Case for Precision Higgs Physics}},  in \emph{{Snowmass
  2021}}, 9, 2022, \href{https://arxiv.org/abs/2209.07510}{{\ttfamily
  2209.07510}}.

\bibitem{Zeng:2015gha}
D.-M. Zeng, S.-Q. Wang, X.-G. Wu and J.-M. Shen, \emph{{The Higgs-boson decay
  $H\;\to \;{gg}$ up to ${\alpha }_{s}^{5}$-order under the minimal momentum
  space subtraction scheme}},
  \href{https://doi.org/10.1088/0954-3899/43/7/075001}{\emph{J. Phys. G}
  {\bfseries 43} (2016) 075001}
  [\href{https://arxiv.org/abs/1507.03222}{{\ttfamily 1507.03222}}].

\bibitem{Zeng:2020lwi}
J.~Zeng, X.-G. Wu, X.-C. Zheng and J.-M. Shen, \emph{{Gauge dependence of the
  perturbative QCD predictions under the momentum space subtraction scheme}},
  \href{https://doi.org/10.1088/1674-1137/abae4e}{\emph{Chin. Phys. C}
  {\bfseries 44} (2020) 113102}
  [\href{https://arxiv.org/abs/2004.12068}{{\ttfamily 2004.12068}}].

\bibitem{Englert:2014aca}
C.~Englert and M.~Spannowsky, \emph{{Limitations and Opportunities of Off-Shell
  Coupling Measurements}},
  \href{https://doi.org/10.1103/PhysRevD.90.053003}{\emph{Phys. Rev. D}
  {\bfseries 90} (2014) 053003}
  [\href{https://arxiv.org/abs/1405.0285}{{\ttfamily 1405.0285}}].

\bibitem{Englert:2014ffa}
C.~Englert, Y.~Soreq and M.~Spannowsky, \emph{{Off-Shell Higgs Coupling
  Measurements in BSM scenarios}},
  \href{https://doi.org/10.1007/JHEP05(2015)145}{\emph{JHEP} {\bfseries 05}
  (2015) 145} [\href{https://arxiv.org/abs/1410.5440}{{\ttfamily 1410.5440}}].

\bibitem{Taylor:1971ff}
J.~C. Taylor, \emph{{Ward Identities and Charge Renormalization of the
  Yang-Mills Field}},
  \href{https://doi.org/10.1016/0550-3213(71)90297-5}{\emph{Nucl. Phys. B}
  {\bfseries 33} (1971) 436}.

\bibitem{Slavnov:1972fg}
A.~A. Slavnov, \emph{{Ward Identities in Gauge Theories}},
  \href{https://doi.org/10.1007/BF01090719}{\emph{Theor. Math. Phys.}
  {\bfseries 10} (1972) 99}.

\bibitem{Kugo:1979gm}
T.~Kugo and I.~Ojima, \emph{{Local Covariant Operator Formalism of Nonabelian
  Gauge Theories and Quark Confinement Problem}},
  \href{https://doi.org/10.1143/PTPS.66.1}{\emph{Prog. Theor. Phys. Suppl.}
  {\bfseries 66} (1979) 1}.

\bibitem{CDF:2022hxs}
{\scshape CDF} collaboration, \emph{{High-precision measurement of the $W$
  boson mass with the CDF II detector}},
  \href{https://doi.org/10.1126/science.abk1781}{\emph{Science} {\bfseries 376}
  (2022) 170}.

\bibitem{ATLAS:2024erm}
{\scshape ATLAS} collaboration, \emph{{Measurement of the $W$-boson mass and
  width with the ATLAS detector using proton-proton collisions at $\sqrt{s}$ =
  7 TeV}},  \href{https://arxiv.org/abs/2403.15085}{{\ttfamily 2403.15085}}.

\bibitem{CMS-PAS-SMP-23-002}
{\scshape CMS} collaboration, \emph{{Measurement of the W boson mass in
  proton-proton collisions at $\sqrt{s} =$ 13 TeV}},  tech. rep., CERN, Geneva,
  2024.

\bibitem{Banerjee:2024eyo}
S.~Banerjee, D.~Reichelt and M.~Spannowsky, \emph{{Electroweak corrections and
  EFT operators in W+W- production at the LHC}},
  \href{https://doi.org/10.1103/PhysRevD.110.115012}{\emph{Phys. Rev. D}
  {\bfseries 110} (2024) 115012}
  [\href{https://arxiv.org/abs/2406.15640}{{\ttfamily 2406.15640}}].

\bibitem{ATLAS:2023gsl}
{\scshape ATLAS} collaboration, \emph{{Inclusive and differential
  cross-sections for dilepton $ t\overline{t} $ production measured in $
  \sqrt{s} $ = 13 TeV pp collisions with the ATLAS detector}},
  \href{https://doi.org/10.1007/JHEP07(2023)141}{\emph{JHEP} {\bfseries 07}
  (2023) 141} [\href{https://arxiv.org/abs/2303.15340}{{\ttfamily
  2303.15340}}].

\bibitem{Buckley:2015lku}
A.~Buckley, C.~Englert, J.~Ferrando, D.~J. Miller, L.~Moore, M.~Russell et~al.,
  \emph{{Constraining top quark effective theory in the LHC Run II era}},
  \href{https://doi.org/10.1007/JHEP04(2016)015}{\emph{JHEP} {\bfseries 04}
  (2016) 015} [\href{https://arxiv.org/abs/1512.03360}{{\ttfamily
  1512.03360}}].

\bibitem{Hartland:2019bjb}
N.~P. Hartland, F.~Maltoni, E.~R. Nocera, J.~Rojo, E.~Slade, E.~Vryonidou
  et~al., \emph{{A Monte Carlo global analysis of the Standard Model Effective
  Field Theory: the top quark sector}},
  \href{https://doi.org/10.1007/JHEP04(2019)100}{\emph{JHEP} {\bfseries 04}
  (2019) 100} [\href{https://arxiv.org/abs/1901.05965}{{\ttfamily
  1901.05965}}].

\bibitem{Brivio:2019ius}
I.~Brivio, S.~Bruggisser, F.~Maltoni, R.~Moutafis, T.~Plehn, E.~Vryonidou
  et~al., \emph{{O new physics, where art thou? A global search in the top
  sector}}, \href{https://doi.org/10.1007/JHEP02(2020)131}{\emph{JHEP}
  {\bfseries 02} (2020) 131}
  [\href{https://arxiv.org/abs/1910.03606}{{\ttfamily 1910.03606}}].

\bibitem{Bissmann:2020mfi}
S.~Bi\ss{}mann, C.~Grunwald, G.~Hiller and K.~Kr\"oninger, \emph{{Top and
  Beauty synergies in SMEFT-fits at present and future colliders}},
  \href{https://doi.org/10.1007/JHEP06(2021)010}{\emph{JHEP} {\bfseries 06}
  (2021) 010} [\href{https://arxiv.org/abs/2012.10456}{{\ttfamily
  2012.10456}}].

\bibitem{Ethier:2021bye}
{\scshape SMEFiT} collaboration, \emph{{Combined SMEFT interpretation of Higgs,
  diboson, and top quark data from the LHC}},
  \href{https://doi.org/10.1007/JHEP11(2021)089}{\emph{JHEP} {\bfseries 11}
  (2021) 089} [\href{https://arxiv.org/abs/2105.00006}{{\ttfamily
  2105.00006}}].

\bibitem{Garosi:2023yxg}
F.~Garosi, D.~Marzocca, A.~R. S\'anchez and A.~Stanzione, \emph{{Indirect
  constraints on top quark operators from a global SMEFT analysis}},
  \href{https://doi.org/10.1007/JHEP12(2023)129}{\emph{JHEP} {\bfseries 12}
  (2023) 129} [\href{https://arxiv.org/abs/2310.00047}{{\ttfamily
  2310.00047}}].

\bibitem{Atkinson:2024hqp}
O.~Atkinson, C.~Englert, M.~Kirk and G.~Tetlalmatzi-Xolocotzi,
  \emph{{Collider-Flavour Complementarity from the bottom to the top}},
  \href{https://arxiv.org/abs/2411.00940}{{\ttfamily 2411.00940}}.

\bibitem{Aguilar-Saavedra:2024mnm}
J.~A. Aguilar-Saavedra, \emph{{Toponium hunter\textquoteright{}s guide}},
  \href{https://doi.org/10.1103/PhysRevD.110.054032}{\emph{Phys. Rev. D}
  {\bfseries 110} (2024) 054032}
  [\href{https://arxiv.org/abs/2407.20330}{{\ttfamily 2407.20330}}].

\bibitem{Fuks:2024yjj}
B.~Fuks, K.~Hagiwara, K.~Ma and Y.-J. Zheng, \emph{{Simulating toponium
  formation signals at the LHC}},
  \href{https://arxiv.org/abs/2411.18962}{{\ttfamily 2411.18962}}.

\bibitem{Garzelli:2024uhe}
M.~V. Garzelli, G.~Limatola, S.~O. Moch, M.~Steinhauser and O.~Zenaiev,
  \emph{{Updated predictions for toponium production at the LHC}},
  \href{https://arxiv.org/abs/2412.16685}{{\ttfamily 2412.16685}}.

\bibitem{Dreiner:2008tw}
H.~K. Dreiner, H.~E. Haber and S.~P. Martin, \emph{{Two-component spinor
  techniques and Feynman rules for quantum field theory and supersymmetry}},
  \href{https://doi.org/10.1016/j.physrep.2010.05.002}{\emph{Phys. Rept.}
  {\bfseries 494} (2010) 1} [\href{https://arxiv.org/abs/0812.1594}{{\ttfamily
  0812.1594}}].

\bibitem{Englert:2018byk}
C.~Englert, M.~Russell and C.~D. White, \emph{{Effective Field Theory in the
  top sector: do multijets help?}},
  \href{https://doi.org/10.1103/PhysRevD.99.035019}{\emph{Phys. Rev. D}
  {\bfseries 99} (2019) 035019}
  [\href{https://arxiv.org/abs/1809.09744}{{\ttfamily 1809.09744}}].

\bibitem{Grzadkowski:2010es}
B.~Grzadkowski, M.~Iskrzynski, M.~Misiak and J.~Rosiek, \emph{{Dimension-Six
  Terms in the Standard Model Lagrangian}},
  \href{https://doi.org/10.1007/JHEP10(2010)085}{\emph{JHEP} {\bfseries 10}
  (2010) 085} [\href{https://arxiv.org/abs/1008.4884}{{\ttfamily 1008.4884}}].

\bibitem{Celada:2024mcf}
E.~Celada, T.~Giani, J.~ter Hoeve, L.~Mantani, J.~Rojo, A.~N. Rossia et~al.,
  \emph{{Mapping the SMEFT at high-energy colliders: from LEP and the (HL-)LHC
  to the FCC-ee}}, \href{https://doi.org/10.1007/JHEP09(2024)091}{\emph{JHEP}
  {\bfseries 09} (2024) 091}
  [\href{https://arxiv.org/abs/2404.12809}{{\ttfamily 2404.12809}}].

\bibitem{ATLAS:2019onj}
{\scshape ATLAS} collaboration, \emph{{Measurement of the top-quark decay width
  in top-quark pair events in the dilepton channel at $\sqrt{s}=13$ TeV with
  the ATLAS detector}},  \href{https://arxiv.org/abs/ATLAS-CONF-2019-038,
  ATLAS-CONF-2019-038}{{\ttfamily ATLAS-CONF-2019-038, ATLAS-CONF-2019-038}}.

\bibitem{FCC:2018byv}
{\scshape FCC} collaboration, \emph{{FCC Physics Opportunities}: {Future
  Circular Collider Conceptual Design Report Volume 1}},
  \href{https://doi.org/10.1140/epjc/s10052-019-6904-3}{\emph{Eur. Phys. J. C}
  {\bfseries 79} (2019) 474}.

\bibitem{FCC:2018evy}
{\scshape FCC} collaboration, \emph{{FCC-ee: The Lepton Collider}: {Future
  Circular Collider Conceptual Design Report Volume 2}},
  \href{https://doi.org/10.1140/epjst/e2019-900045-4}{\emph{Eur. Phys. J. ST}
  {\bfseries 228} (2019) 261}.

\bibitem{ILC:2013jhg}
{\scshape ILC} collaboration, \emph{{The International Linear Collider
  Technical Design Report - Volume 2: Physics}},
  \href{https://arxiv.org/abs/1306.6352}{{\ttfamily 1306.6352}}.

\bibitem{CEPCStudyGroup:2018rmc}
{\scshape CEPC Study Group} collaboration, \emph{{CEPC Conceptual Design
  Report: Volume 1 - Accelerator}},
  \href{https://arxiv.org/abs/1809.00285}{{\ttfamily 1809.00285}}.

\bibitem{Lebrun:2012hj}
P.~Lebrun, L.~Linssen, A.~Lucaci-Timoce, D.~Schulte, F.~Simon, S.~Stapnes
  et~al., \emph{{The CLIC Programme: Towards a Staged e+e- Linear Collider
  Exploring the Terascale : CLIC Conceptual Design Report}},
  \href{https://arxiv.org/abs/1209.2543}{{\ttfamily 1209.2543}}.

\bibitem{CLICdp:2018cto}
{\scshape CLICdp, CLIC} collaboration, \emph{{The Compact Linear Collider
  (CLIC) - 2018 Summary Report}},
  \href{https://arxiv.org/abs/1812.06018}{{\ttfamily 1812.06018}}.

\bibitem{Manosperti:2023sbd}
E.~Manosperti, R.~Tomas and A.~Pastushenko, \emph{{Design of CLIC beam delivery
  system at 7 TeV}},
  \href{https://doi.org/10.18429/JACoW-IPAC2023-MOPL113}{\emph{JACoW}
  {\bfseries IPAC2023} (2023) MOPL113}.

\bibitem{AlAli:2021let}
H.~Al~Ali et~al., \emph{{The muon Smasher\textquoteright{}s guide}},
  \href{https://doi.org/10.1088/1361-6633/ac6678}{\emph{Rept. Prog. Phys.}
  {\bfseries 85} (2022) 084201}
  [\href{https://arxiv.org/abs/2103.14043}{{\ttfamily 2103.14043}}].

\bibitem{Chiesa:2020awd}
M.~Chiesa, F.~Maltoni, L.~Mantani, B.~Mele, F.~Piccinini and X.~Zhao,
  \emph{{Measuring the quartic Higgs self-coupling at a multi-TeV muon
  collider}}, \href{https://doi.org/10.1007/JHEP09(2020)098}{\emph{JHEP}
  {\bfseries 09} (2020) 098}
  [\href{https://arxiv.org/abs/2003.13628}{{\ttfamily 2003.13628}}].

\bibitem{Costantini:2020stv}
A.~Costantini, F.~De~Lillo, F.~Maltoni, L.~Mantani, O.~Mattelaer, R.~Ruiz
  et~al., \emph{{Vector boson fusion at multi-TeV muon colliders}},
  \href{https://doi.org/10.1007/JHEP09(2020)080}{\emph{JHEP} {\bfseries 09}
  (2020) 080} [\href{https://arxiv.org/abs/2005.10289}{{\ttfamily
  2005.10289}}].

\bibitem{Celada:2023oji}
E.~Celada, T.~Han, W.~Kilian, N.~Kreher, Y.~Ma, F.~Maltoni et~al.,
  \emph{{Probing Higgs-muon interactions at a multi-TeV muon collider}},
  \href{https://doi.org/10.1007/JHEP08(2024)021}{\emph{JHEP} {\bfseries 08}
  (2024) 021} [\href{https://arxiv.org/abs/2312.13082}{{\ttfamily
  2312.13082}}].

\bibitem{Dedes:2017zog}
A.~Dedes, W.~Materkowska, M.~Paraskevas, J.~Rosiek and K.~Suxho, \emph{{Feynman
  rules for the Standard Model Effective Field Theory in R$_{\xi}$-gauges}},
  \href{https://doi.org/10.1007/JHEP06(2017)143}{\emph{JHEP} {\bfseries 06}
  (2017) 143} [\href{https://arxiv.org/abs/1704.03888}{{\ttfamily
  1704.03888}}].

\bibitem{Dedes:2019uzs}
A.~Dedes, M.~Paraskevas, J.~Rosiek, K.~Suxho and L.~Trifyllis, \emph{{SmeftFR
  \textendash{} Feynman rules generator for the Standard Model Effective Field
  Theory}}, \href{https://doi.org/10.1016/j.cpc.2019.106931}{\emph{Comput.
  Phys. Commun.} {\bfseries 247} (2020) 106931}
  [\href{https://arxiv.org/abs/1904.03204}{{\ttfamily 1904.03204}}].

\bibitem{Dedes:2023zws}
A.~Dedes, J.~Rosiek, M.~Ryczkowski, K.~Suxho and L.~Trifyllis, \emph{{SmeftFR
  v3 \textendash{} Feynman rules generator for the Standard Model Effective
  Field Theory}},
  \href{https://doi.org/10.1016/j.cpc.2023.108943}{\emph{Comput. Phys. Commun.}
  {\bfseries 294} (2024) 108943}
  [\href{https://arxiv.org/abs/2302.01353}{{\ttfamily 2302.01353}}].

\bibitem{Alloul:2013bka}
A.~Alloul, N.~D. Christensen, C.~Degrande, C.~Duhr and B.~Fuks,
  \emph{{FeynRules 2.0 - A complete toolbox for tree-level phenomenology}},
  \href{https://doi.org/10.1016/j.cpc.2014.04.012}{\emph{Comput. Phys. Commun.}
  {\bfseries 185} (2014) 2250}
  [\href{https://arxiv.org/abs/1310.1921}{{\ttfamily 1310.1921}}].

\bibitem{Degrande:2011ua}
C.~Degrande, C.~Duhr, B.~Fuks, D.~Grellscheid, O.~Mattelaer and T.~Reiter,
  \emph{{UFO - The Universal FeynRules Output}},
  \href{https://doi.org/10.1016/j.cpc.2012.01.022}{\emph{Comput. Phys. Commun.}
  {\bfseries 183} (2012) 1201}
  [\href{https://arxiv.org/abs/1108.2040}{{\ttfamily 1108.2040}}].

\bibitem{Alwall:2011uj}
J.~Alwall, M.~Herquet, F.~Maltoni, O.~Mattelaer and T.~Stelzer, \emph{{MadGraph
  5: Going Beyond}}, \href{https://doi.org/10.1007/JHEP06(2011)128}{\emph{JHEP}
  {\bfseries 06} (2011) 128} [\href{https://arxiv.org/abs/1106.0522}{{\ttfamily
  1106.0522}}].

\bibitem{Murayama:1992gi}
H.~Murayama, I.~Watanabe and K.~Hagiwara, \emph{{HELAS: HELicity amplitude
  subroutines for Feynman diagram evaluations}},
  \href{https://arxiv.org/abs/KEK-91-11}{{\ttfamily KEK-91-11}}.

\bibitem{Seidel:2013sqa}
K.~Seidel, F.~Simon, M.~Tesar and S.~Poss, \emph{{Top quark mass measurements
  at and above threshold at CLIC}},
  \href{https://doi.org/10.1140/epjc/s10052-013-2530-7}{\emph{Eur. Phys. J. C}
  {\bfseries 73} (2013) 2530}
  [\href{https://arxiv.org/abs/1303.3758}{{\ttfamily 1303.3758}}].

\bibitem{Chekanov:2017tfl}
S.~Chekanov, M.~Demarteau, A.~Fischer and J.~Zhang, \emph{{Effect of PYTHIA8
  tunes on event shapes and top-quark reconstruction in $e^+e^-$ annihilation
  at CLIC}},  \href{https://arxiv.org/abs/1710.07713}{{\ttfamily 1710.07713}}.

\bibitem{Li:2022iav}
Z.~Li, X.~Sun, Y.~Fang, G.~Li, S.~Xin, S.~Wang et~al., \emph{{Top quark mass
  measurements at the $t\bar{t}$ threshold with CEPC}},
  \href{https://doi.org/10.1140/epjc/s10052-023-11421-1}{\emph{Eur. Phys. J. C}
  {\bfseries 83} (2023) 269}
  [\href{https://arxiv.org/abs/2207.12177}{{\ttfamily 2207.12177}}].

\bibitem{Durieux:2018tev}
G.~Durieux, M.~Perell\'o, M.~Vos and C.~Zhang, \emph{{Global and optimal probes
  for the top-quark effective field theory at future lepton colliders}},
  \href{https://doi.org/10.1007/JHEP10(2018)168}{\emph{JHEP} {\bfseries 10}
  (2018) 168} [\href{https://arxiv.org/abs/1807.02121}{{\ttfamily
  1807.02121}}].

\bibitem{CLICdp:2018esa}
{\scshape CLICdp} collaboration, \emph{{Top-Quark Physics at the CLIC
  Electron-Positron Linear Collider}},
  \href{https://doi.org/10.1007/JHEP11(2019)003}{\emph{JHEP} {\bfseries 11}
  (2019) 003} [\href{https://arxiv.org/abs/1807.02441}{{\ttfamily
  1807.02441}}].

\bibitem{Englert:2017dev}
C.~Englert and M.~Russell, \emph{{Top quark electroweak couplings at future
  lepton colliders}},
  \href{https://doi.org/10.1140/epjc/s10052-017-5095-z}{\emph{Eur. Phys. J. C}
  {\bfseries 77} (2017) 535}
  [\href{https://arxiv.org/abs/1704.01782}{{\ttfamily 1704.01782}}].

\end{thebibliography}\endgroup

\end{document}